\documentclass[sigconf]{acmart}

\AtBeginDocument{%
  }

\usepackage{color}
\usepackage{amsmath}
\usepackage{ascmac}
\usepackage{amsthm}

\usepackage{booktabs} 
\usepackage{subcaption}

\setcopyright{none}
\copyrightyear{}
\acmYear{}
\acmDOI{}
\acmConference[]{}{}{}
\acmISBN{}
\renewcommand\footnotetextcopyrightpermission[1]{}

\begin{document}
\pagestyle{empty}

\title{Do Expressions Change Decisions? Exploring the Impact of AI's Explanation Tone on Decision-Making}

%
\author{Ayano Okoso}
\affiliation{%
  \institution{Toyota Central R\&D Labs., Inc. }
  \streetaddress{41-1 Yokomichi Nagakute}
  \city{Aichi}
  \country{Japan}}
\email{okoso@mosk.tytlabs.co.jp}

\author{Mingzhe Yang}
\affiliation{%
  \institution{The University of Tokyo}
  \streetaddress{3-8-1 Komaba Meguro}
  \city{Tokyo}
  \country{Japan}}
\email{mingzhe-yang@g.ecc.u-tokyo.ac.jp}

\author{Yukino Baba}
\affiliation{%
  \institution{The University of Tokyo}
  \streetaddress{3-8-1 Komaba Meguro}
  \city{Tokyo}
  \country{Japan}}
\email{yukino-baba@g.ecc.u-tokyo.ac.jp}

%
\renewcommand{\shortauthors}{okoso et al.}

\begin{abstract}
Explanatory information helps users to evaluate the suggestions offered by AI-driven decision support systems.
With large language models, adjusting explanation expressions has become much easier.
However, how these expressions influence human decision-making remains largely unexplored.
This study investigated the effect of explanation tone (e.g., formal or humorous) on decision-making, focusing on AI roles and user attributes.
We conducted user experiments across three scenarios depending on AI roles (assistant, second-opinion provider, and expert) using datasets designed with varying tones.
The results revealed that tone significantly influenced decision-making regardless of user attributes in the second-opinion scenario, whereas its impact varied by user attributes in the assistant and expert scenarios. 
In addition, older users were more influenced by tone, and highly extroverted users exhibited discrepancies between their perceptions and decisions.
Furthermore, open-ended questionnaires highlighted that users expect tone adjustments to enhance their experience while emphasizing the importance of tone consistency and ethical considerations. Our findings provide crucial insights  into the design of explanation expressions.
\end{abstract}

\ccsdesc[500]{Human-centered computing~User studies}
\ccsdesc[500]{Information systems~Recommender systems}
\ccsdesc[500]{Information systems~Decision support systems}
\ccsdesc[300]{Computing methodologies~Natural language generation}
\keywords{Decision-making Support System, Recommender System, Explanation, Expression, Tone, Large Language Model}


\maketitle

\section{Introduction}\label{sec:intro}

Artificial intelligence (AI) is increasingly important in human decision-making and has become deeply embedded in our daily lives. 
AI is used in many contexts, such as recommender systems on e-commerce platforms~\cite{resnick1997recommender} and search engine algorithms~\cite{joachims2002optimizing}, and in specialized areas, such as tools for predicting recidivism risk~\cite{angwin2022machine,grgic2019human,dressel2018accuracy} and medical diagnosis~\cite{cai2019human,tsai2021exploring}.
As these systems become more complex and influential, it is crucial not only to improve their accuracy but also to understand how they communicate their decisions to users and how this communication impacts users' decision-making.

A key element of this communication is explanations provided by AI~\cite{zhang2020explainable,dovsilovic2018explainable,christoph2020interpretable}.
Explanations are intended to assist decision-making by presenting the model's logic, the basis for its predictions, justifications for the results, additional information, and the training data used~\cite{zhang2020explainable,lai2023towards}.
For example, in a loan default prediction task~\cite{lai2023towards}, an explanation like ``This individual is likely to default because their credit score is poor and their annual income is under \$40,000'' presents both the prediction and the key factors behind it.
In a recommendation context, an explanation such as ``We recommended this romance movie based on your viewing history'' justifies the system's suggestion.
These explanations clarify the reasoning behind the AI's decisions, enhance user understanding~\cite{guo2019visualizing,cai2019effects}, and build trust~\cite{ma2023should,zhang2020effect}.

In human-to-human communication, not only the content of explanations but also how they are delivered---\emph{expressions}---significantly influence the recipient's decision-making process~\cite{bonvillain2019language,scherer2003vocal}.
For instance, a friendly or empathetic tone may make advice more persuasive, whereas a condescending tone may lead to the rejection of sound advice. 
Given this understanding of human communication, an important question arises: \emph{Do the expressions of AI-generated explanations similarly influence human decision-making?}

Understanding how the expressions of AI explanations influence user decision-making is crucial for developing reliable decision support systems.
With the rise of large language models (LLMs), adjusting expressions, such as the tone and style, of explanations has become much easier, but can have both positive and negative effects.
An AI may suggest a product that does not align with the user's preferences, yet how the explanation is conveyed can capture the user's interests, leading to acceptance.
If a user enjoys the product, it might be considered a serendipitous discovery.
However, if not, the user may feel misled, causing a loss of trust.
In other decision-support contexts, unpleasant expressions may cause users to react negatively and decide against their original intentions.
This could lead to excessive trust in the system or disregard for important advice, both of which affect the system's reliability.
Therefore, \emph{our research purpose is to uncover and understand the impact of the expression of AI explanations on decision-making.}

While the content of AI explanations is found to influence decision-making~\cite{gedikli2014should,oyebode2021tailoring}, research on the expressions has primarily explored how they affect user perceptions of AI systems~\cite{jung2022great,okoso2024impact}, leaving their impact on decision-making largely unexplored.
The gap between perception and decision-making cannot be easily bridged by the simplistic assumption that behavior follows perception.
Some studies have found that users may assert trust in an AI system yet ignore its suggestions~\cite{Schaffer2019iui,rechkemmer2022confidence,papenmeier2022s,rong2023towards}.
Possible reasons include the lack of a one-to-one correspondence between perception and behavior, user overconfidence (e.g.,``The AI performs well, but I can do better''), and biases such as politeness bias or the observer-expectancy effect. 
These gaps highlight the importance of investigating the impact of explanation expressions on user decision-making.

Although AI systems serve various roles~\cite{bader1988practical}, does the impact of explanation expressions on decision-making remain the same across these roles?
Users' expectations for explanations vary depending on AI roles~\cite{arrieta2020explainable,ferreira2021human}, which suggests that the impact of explanation expressions on decision-making may also differ depending on AI roles.
Bader et al.~\cite{bader1988practical} identified six roles for knowledge-based systems: assistant, critic, second-opinion provider, expert, teacher, and automation.
For instance, in scenarios requiring specialized expertise, such as legal or medical contexts, AI systems often act as experts or second-opinion providers.
In such cases, users may prioritize the content of their explanations, preferring straightforward and simple expressions.
In contrast, in highly subjective scenarios such as entertainment, AI systems are often expected to serve as assistants, where explanation expressions are as important as the content, and some users may prefer richer and more engaging expressions.
These considerations led us to the following research question: \emph{Does the impact of explanation expressions on decision-making vary depending on the role of the AI system?}

The impact of explanation expressions on decision-making can vary depending on the user as well as the AI roles.
Previous studies have actually shown that the influence of explanation expressions on perceptions of an AI system can differ based on user attributes, such as age and personality traits~\cite{okoso2024impact,chin2024like}.
If the impact of explanation expressions also varies on decision-making depending on user attributes, as it does on perceptions, there is a risk of unintentionally introducing biases or excessively influencing certain user groups.
To build an ideal AI system that provides explanations useful for a diverse range of users, it is crucial to understand how the effects of explanation expressions vary across different user attributes.
Hence, we pose the following question: \emph{How does the impact of explanation expressions on decision-making vary across user attributes?}

Based on the above discussion, this study addressed the following specific research questions.
Following previous studies~\cite{umap2024okoso, okoso2024impact, chin2024like}, we adopted \emph{tones} (e.g., humorous, formal) shown to have varying effects on user perceptions of AI systems, depending on user attributes.
\begin{itemize}
\item \textbf{RQ1:} Does the tone of explanation reinforce or contradict users' initial intentions, thereby influencing their decision-making?
\item \textbf{RQ2:} How does the impact of explanation tone on decision-making vary across different AI roles?
\item \textbf{RQ3:} How does the impact of explanation tone on decision-making vary depending on user attributes?
\end{itemize}

To address these research questions, we conducted user experiments using a dataset generated with various explanation tones.
Assuming a connection between risk levels and the roles of AI systems, we selected three scenarios, each characterized by different risk levels: \emph{movie recommendations}, \emph{opinion formation on controversial topics}, and \emph{advice on recidivism risk prediction tasks}.
In these scenarios, we assumed that the AI system functioned as an \emph{assistant}, \emph{second-opinion provider}, and \emph{expert}, respectively, with the latter representing higher-risk scenarios.
Based on previous research~\cite{okoso2024impact}, we examined the relationship between explanation tones and user attributes by focusing on age, gender, and the Big Five personality traits~\cite{john1999big}.
Furthermore, we conducted a questionnaire-based survey to investigate user expectations and concerns regarding AI-driven tone adjustment.

Our study reveals that the tone of explanations can influence user decision-making and provides insights into understanding these effects from the perspectives of AI roles and user attributes. 
We summarize our main contributions as follows:
\begin{itemize}
    \item This study demonstrated that the impact of explanation tones on decision-making varies depending on AI roles. Specifically, in the opinion formation scenario where the AI served as a second-opinion provider, the tone of explanation significantly influenced decision-making regardless of user attributes. In contrast, in scenarios where the AI acted as an assistant or expert, such as movie recommendations or recidivism risk prediction, the effect of explanation tones on decision-making differed based on user attributes.
    \item This study showed that the impact of explanation tones on decision-making varies depending on user attributes.
    Our findings include that older users were more susceptible to the influence of explanation tones on their decision-making, and the same explanation tone had different effects on decision-making depending on user attributes.
    \item Through open-ended survey responses, this study identified users' expectations and concerns regarding AI-driven tone adjustment. Users expressed expectations for tone adjustment to improve clarity, empathy, and the presentation of diverse perspectives. However, they also raised concerns about the potential for manipulation or bias, and confusion caused by inconsistent tone. We discussed practical approaches to address these concerns.
\end{itemize}

The remainder of this paper is structured as follows. Section~\ref{sec:rw} reviews related work. Section~\ref{sec:overview} provides an overview of the user experiments. Sections~\ref{sec:recsys} --~\ref{sec:advice} detail the user experiments for each scenario. 
Section~\ref{sec:discussion} discusses the interpretations and generalizability of the results, user concerns, countermeasures regarding tone adjustments, and the limitations of our study.
Finally, Section~\ref{sec:conclusion} presents the conclusions.

\section{Related Work}\label{sec:rw}

\subsection{Impact of AI Explanation Content on Decision-Making}\label{subsec:rw1}

With the spread of AI, decision-making support systems are increasingly being used in various fields such as e-commerce~\cite{schafer2001commerce,linden2003amazon}, law~\cite{angwin2022machine,grgic2019human,dressel2018accuracy}, medicine~\cite{cai2019human,tsai2021exploring}, and finance~\cite{hase2020evaluating,yang2024fair,binns2018s}.
Specifically in high-risk areas, where potential for significant losses remains, full automation is often undesirable. 
AI systems commonly provide predictions or advice for decision-making tasks; however, the final decision is left to humans.
Additionally, for recommendation systems, providing explanations for suggested items is crucial for helping users understand and trust the recommendations~\cite{zhang2020explainable,cramer2008effects}.

According to the literature~\cite{lai2023towards,miller2019explanation,zhang2020explainable}, the information presented is not limited to predictions or suggestions, but can be categorized into prediction- and model-related information.
Prediction-related information includes uncertainty of the predictions~\cite{guo2019visualizing,lee2021human}, local feature importance~\cite{alqaraawi2020evaluating,chromik2021think}, rule-based explanations~\cite{lim2009and,lakkaraju2016interpretable}, and example-based explanations~\cite{binns2018s,cai2019human}.
Model-related information includes model accuracy~\cite{lai2020chicago,yang2020visual}, global feature importance~\cite{dodge2019explaining,wang2021explanations}, and the data used for training~\cite{bhattacharya2024exmos}.
Providing these types of information improves task performance and efficiency, as well as the understanding and trust of users in AI systems~\cite{lai2023towards,miller2019explanation,zhang2020explainable}.

Various studies have demonstrated the impact of explanatory content on decision-making.
Gedikli et al.~\cite{gedikli2014should} demonstrated that including elements such as authority and social proof of explanations for movie recommendations can increase consumption intentions for items initially perceived as less appealing.
Similarly, Oyebode et al.~\cite{oyebode2021tailoring} found that for smoking cessation, different types of content are effective depending on the stage of behavioral change.
In decision-making tasks involving AI-provided advice, the content of explanations influence the trust of users in the system and its advice, which affects their acceptance of the advice.
At the same time, the effectiveness of explanation content depends on factors such as the accuracy of the AI system~\cite{kahr2023seems}, user’s domain knowledge~\cite{wang2021explanations}, data format of the task~\cite{hase2020evaluating}, and presentation format of the explanation.~\cite{herlocker2000explaining,kouki2019personalized}.
Moreover, incorrect explanations can induce trust in high-difficulty tasks, leading users to follow inaccurate advice~\cite{sadeghi2024explaining}.
However, perceived trust in AI systems or their advice does not always translate into users accepting the advice.
This discrepancy is more likely when users have extensive domain knowledge, the scenario involves high-risk decision-making~\cite{kim2023humans}, or users report high familiarity with the task~\cite{Schaffer2019iui}.

\subsection{Impact of AI Explanation Expression on Decision-Making}\label{subsec:rw2}

Previous research has explored the impact of various expression styles in AI explanations, particularly focusing on warmth, metaphor, anthropomorphism, and tone, and how they influence user perceptions and behavior.
Several studies have investigated the relationship between warmth and competence in explanations.
Gilad et al.~\cite{gilad2021effects} showed that warmth tends to be prioritized over competence in AI system selection.
Similarly, Khadpe et al.~\cite{khadpe2020conceptual} found that in cooperative tasks, AI agents perceived as high in warmth but low in competence increased the ease of use, intention to use, and willingness to cooperate.
Mckee et al.~\cite{mckee2024warmth} demonstrated that users’ perceptions of agents are influenced more by warmth and competence than by objective performance.
Additionally, a literature review~\cite{jung2023female} indicated that female chatbots are more likely to evoke warmth and empathy, suggesting the importance of constructing metaphors that emphasize warmth rather than relying on binary gender distinctions.
In the context of home healthcare triage, G{\'o}mez et al.~\cite{gomez4797707large} found that the effect of the profile of an agent (rational vs. empathetic) on its advice was insignificant.

Regarding other methods of expression, Jung et al.~\cite{jung2022great} examined the impact of conceptual metaphors on crowdworkers in conversational agents.
They found that animal metaphors elicited higher engagement than inanimate metaphors, although no significant differences were noted in trust across metaphors.
Yang et al.~\cite{yang2024role} found that users react negatively when AI recommends vice products, but anthropomorphizing AI mitigates these negative reactions.
On voice assistants for older users, Chin et al.~\cite{chin2024like} found that those with lower agreeableness preferred informal conversational styles.
Additionally, in recommendation systems, Okoso et al.~\cite{okoso2024impact} demonstrated that the tone of explanations (formal or humorous) affects users’ perceptions of trustworthiness and persuasiveness, with these effects varying based on user attributes.
Moreover, several studies demonstrated that tailoring advertisements using LLMs to match the personality traits of users increases the likelihood of persuasion~\cite{meguellati2024good,matz2024potential}.

\subsection{Research Gaps}\label{subsec:gap}

As previously mentioned, extensive studies have been conducted on the explanations provided by AI systems in various decision-making support scenarios.
These studies have primarily focused on the types of explanation content that are effective and how they influence task performance and trust in the system, often discussing explanation content and expression separately.
Although studies focusing on explanation expressions have been increasing, most of these studies emphasize user perceptions, such as trust and understandability of the system, rather than their actual impact on decision-making.

Interestingly, perception and decision-making are not necessarily aligned; even when users claim to trust or understand an AI system, they do not always follow its suggestions~\cite{Schaffer2019iui,rechkemmer2022confidence,papenmeier2022s,wang2021explanations,hase2020evaluating}.
Whether these discrepancies also occur in explanation expressions remains unexplored.
Furthermore, existing studies on explanation expressions have mainly been limited to specific scenarios~\cite{yang2024fair,sivaraman2023ignore,bansal2021does,kahr2023seems,wang2021explanations}, leaving a lack of discussion on how these expressions influence decision-making across scenarios with different AI system roles.
Similarly, studies addressing user attributes remain limited~\cite{okoso2024impact,chin2024like}, and whether the impact of explanation expressions differs among user groups with varying attributes remains insufficiently understood.
To address these gaps, our study aims to investigate the impact of explanation expressions on decision-making by considering multiple scenarios with varying AI system roles and incorporating user attributes.

\begin{table*}[ht]
\centering
\small
\caption{Definitions of each tone}
\label{tb:tone_definitions}
\begin{tabular}{ll}
\toprule
\textbf{Tone} & \textbf{Definition} \\
\midrule
neutral & Objective and factual expression with a restrained emotional tone. Simple and direct language. \\
formal & Polite and sophisticated language. Professional and official tone. Courteous and cautious expression. \\
authoritative & Confident and assertive language. Persuasive and commanding expression. \\
casual & Relaxed and informal language. Friendly and conversational tone, often using colloquial expressions. \\
humorous & Incorporates humor and light-hearted expression. Casual tone with playful language. \\
romantic & Passionate and emotionally rich expression. Poetic and beautiful language. \\
\bottomrule
\end{tabular}
\end{table*}

\section{Overview of User Experiments}\label{sec:overview}

To address our research questions and examine the impact of explanation tone on user decision-making, we conducted three online user experiments. 
We designed three scenarios, each representing different AI roles: recommendation (assistant), opinion formation (second-opinion provider), and advice (expert).
This section provides an overview of each experimental scenario, common procedure, and participants information.

\subsection{Overview of Experiment Scenarios}\label{subsec:overview_scenario}

In each scenario, we assessed the effect of explanation tone by comparing task scores when explanations were presented in a neutral tone versus other tones.
We outline the tasks and explanations for each scenario below.

\paragraph{Recommendation}
The AI acts as an assistant, supporting users in everyday decision-making.
The explanations provided help users determine whether the recommendations align with their preferences.
In this scenario, participants were presented with one movie per task and asked to evaluate how much they would like to watch it.
Movie information included the title, genre, a brief plot, and a poster image.
Additionally, an AI-generated advertisement encouraging the participant to watch the movie was provided as an explanation.
This advertisement helped users assess and rate their interest in the movie.
This task was repeated, and the influence of tone on decision-making was analyzed by measuring changes in movie scores resulting from variations in the tone of the explanation.

\paragraph{Opinion Formation}
The AI acts as a second-opinion provider, providing explanations to reinforce users' opinions or introduce overlooked viewpoints.
In this scenario, participants were presented with a controversial topic in each task and asked to rate their level of agreement or disagreement.
They also rated their confidence in their responses and their knowledge of the topic.
A controversial topic refers to subjects for which a debate is ongoing among people with different views and in which no single, straightforward solution exists.
Examples include \textit{Should children be assigned homework?} or \textit{Is it acceptable to use single-use plastics?}
In this scenario, AI-generated explanations supporting the pro side of the topic were used.
People often rely on Web search engines when seeking information on controversial topics~\cite{SALMERON20132161}.
Based on the information they find, they might reinforce or reconsider their original stance.
By reading the explanations provided, participants gained more information on the topic, helping them form or express their stance.
To avoid any bias based on the position presented in the explanations, all the participants were shown opinions that supported the pro side of the topic.
By repeating this task, we analyzed how changes in tone affect the agreement, confidence, and knowledge scores of the participants, thereby measuring the impact of tone  on decision-making.

\paragraph{Advice}
The AI acts as an expert and provides essential supplemental information to assist in completing the task.
In this scenario, participants were tasked with predicting a defendant’s likelihood of reoffending within two years based on their profile.
They also rated their confidence in their prediction.
The profile included factors such as age, gender, offense, and prior convictions, similar to those used by judges, probation officers, which are used by judges, probation officers, and parole officers to score the likelihood of reoffending across various states~\cite{dressel2018accuracy}.
In this scenario, the explanation consisted of an AI-generated risk prediction score and the rationale behind it based on the profile of the defendant.
By reading the explanation, participants were able to understand the factors to focus on for prediction, and the provided score served as a useful reference point for making more accurate predictions.
This task was repeated multiple times, and by analyzing how prediction and confidence scores changed with variations in tone, we measured the impact of explanation tone on decision-making.

\subsection{Overview of Experiment Procedures}\label{subsec:overview_procedures}

The participants were randomly assigned to a specific tone--referred to as the \emph{intervention tone}--from a set of pre-prepared tones.
The tones were selected based on existing studies~\cite{umap2024okoso}, with different sets customized for each scenario.
Table~\ref{tb:tone_definitions} lists the types of tones used in the experiments and their definitions.
In the user experiments, we compared the outcomes of tasks in which explanations were presented in a neutral tone with those in which explanations were presented in the intervention tone.

\begin{figure*}[t]
    \centering
    \includegraphics[width=0.98\linewidth]{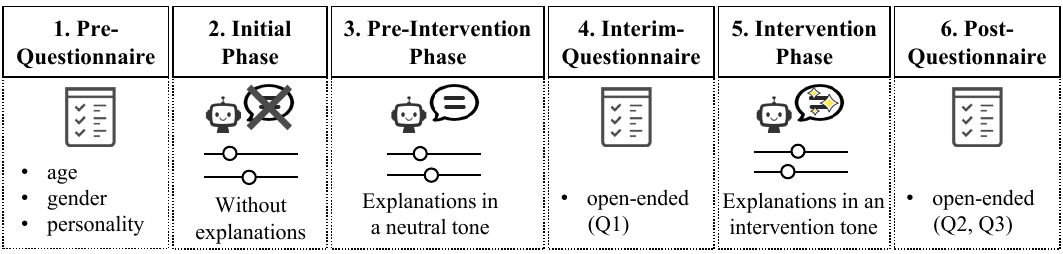}
    \caption[overview]{User Study Procedure. We conducted user studies consisting of six steps. Step 1: participants complete a pre-questionnaire to assess their demographics (age and gender) and Big Five personality traits. Step 2: participants are asked to perform tasks without any presentation of explanations, referred to as the \emph{initial phase}. Step 3: participants complete the tasks with explanations in a neutral tone, referred to as the \emph{pre-intervention phase}. Step 4: participants answer an open-ended questions to assess how much the explanations affected their tasks. Step 5: participants perform the tasks with explanations in an assigned intervention tone, referred to as the \emph{intervention phase}. Step 6: participants are asked to answer open-ended questions to assess how much intervening explanations affect the tasks. The tasks presented to participants were identical across all three phases, ensuring comparability of responses.}
    \Description{The figure shows the user study procedure, consisting of six steps. Step 1: Participants complete a pre-questionnaire to assess their demographics (age and gender) and Big Five personality traits. Step 2: Participants are asked to perform tasks without any explanations, referred to as the ``initial phase.'' Step 3: Participants complete tasks with explanations presented in a neutral tone, called the ``pre-intervention phase.'' Step 4: Participants answer open-ended questions to assess how much the explanations affect their tasks. Step 5: Participants perform tasks with explanations presented in the assigned intervention tone, known as the ``intervention phase.'' Step 6: Participants answer open-ended questions to assess how the intervention explanations affected their tasks. The tasks presented to participants were consistent across all three phases, ensuring comparability of responses.}
    \label{fig:overview}
\end{figure*}

\begin{table*}[t]
\centering
\small
\caption{Open-ended questionnaires}
\label{tb:questionnaire}
\begin{tabular}{lp{13.5cm}}
\toprule
\textbf{ID} & \textbf{Question} \\
\midrule
Q1 & Phases 1 through 2, AI-generated explanations were added. How did this affect your decision-making?\\
Q2 & How did the change in expression affect your decision-making? (From Phase 2 to Phase 3) \\
Q3 & In tasks like the one you just completed, where the system provides explanations to assist in decision-making, how do you feel about AI adjusting the style of these explanations? (expectations, concerns, opinions, or impressions) \\
\bottomrule
\end{tabular}
\end{table*}

Although the tasks differed between the scenarios, the overall procedure of the user experiments was consistent, as shown in Fig.~\ref{fig:overview}.
First, the participants answered questions about their demographics (age and gender) and completed a brief test~\cite{gosling2003very} to assess the Big Five personality traits~\cite{john1999big}.
They then performed the \emph{initial phase (Phase 1)} without any presentation of explanations.
Next, they completed the \emph{pre-intervention phase (Phase 2)} with explanations presented in a neutral tone.
After the pre-intervention phase, they completed an open-ended questionnaire to assess the extent to which the explanations affected the tasks, as shown in Table~\ref{tb:questionnaire} (Q1).
Next, they performed the \emph{intervention phase (Phase 3)} in which explanations were presented in the assigned intervention tone.
Finally, they answered two open-ended questions as shown in Table~\ref{tb:questionnaire} (Q2 and Q3).
While the initial, pre-intervention, and intervention phases differed in the presence or tone of the explanations, the tasks were identical.
Specifically, in the movie recommendation scenario, the participants evaluated the same set of movies; in the opinion formation scenario, they responded to the same set of topics; and in the advice scenario, they assessed the same set of defendants.

Rather than randomly presenting neutral and intervention tones, we divided them into phases and presented the neutral tone first to collect user feedback in a timely manner and clearly evaluate the impact of the intervention tone.
By separating the phases, the participants were able to focus on the differences between having explanations or not, as well as the differences in tones, when answering the questionnaires. 
Additionally, establishing a consistent baseline across participants allowed for a straightforward interpretation of the results.
While potential risks of order and learning effects remained, we mitigated these risks by randomizing both the tasks and their order across participants.

We observed the impact of the presence of explanations by comparing the results of the initial and pre-intervention phases.
Similarly, we observed the impact of the explanation tone by comparing the results of the pre-intervention and intervention phases.
This experimental procedure allowed us to observe how the explanation tone influenced changes in the initial intentions of users.
Notably, the content of the explanations presented in the pre- and intervention phases were identical, with only the tone being controlled.

\subsection{Participants}\label{subsec:participants}

All experiments in this study were conducted as human intelligence tasks (HITs) on Amazon Mechanical Turk (MTurk).
The participants were allowed to respond to only one of the three scenarios.
Because all experiments were conducted in English, participation was restricted to workers residing in English-speaking countries (the United States, United Kingdom, Canada, Ireland, New Zealand, and Australia).
To ensure the quality of participants, the requirement was that MTurk workers had a minimum of $50$ approved HITs and an acceptance rate of $97\%$ or higher.
Moreover, data from participants who did not pass the attention check and those with low-quality responses to the open-ended questions were excluded.
The attention check was embedded in the pre-intervention task, in which a specific instruction was inserted into the explanation, directing the participants to choose a particular option.
Participants were not permitted to use internet searches, translation tools, or generative AI during the experiment.
Each experiment was expected to take $20$--$30$ min, and the participants received $2.50$ USD as compensation.

Table~\ref{tb:participants} lists the number of valid participants assigned to each intervention tone for each scenario.
A total of 292, 338, and 215 participants were present in the recommendation, opinion formation, and advice scenarios, respectively. 
Table~\ref{tb:participants_info} lists the average age and mean values of the personality traits of the participants in each scenario.
The personality traits scores were normalized between 0 and 1 based on the possible score range.
No significant deviation was observed in the distribution of the age or personality traits of the participants  across scenarios.

\begin{table*}[t]
\centering
\small
\caption{Number of participants by intervention tone group}
\label{tb:participants}
\begin{tabular}{lcccccc}
    \toprule
    Scenario & Total & Neutral & Formal & Authoritative & Humorous & Romantic/Casual \\
    \midrule
    Recommendation (Sec.~\ref{sec:recsys}) & 292 (M:140, F:152) & 59 & 74 & 54 & 47 & 58 \\
    Opinion Formation (Sec.~\ref{sec:opinion}) & 338 (M:169, F:169) & 73 & 65 & 72 & 62 & 66 \\
    Advice (Sec.~\ref{sec:advice}) & 215 (M:93, F:122) & 59 & 37 & 39 & 45 & 35 \\
    \bottomrule
\end{tabular}
\end{table*}

\begin{table*}[t]
\centering
\small
\caption{Average age and mean values of personality traits (normalized between 0 and 1) for participants in each scenario. Values in parentheses represent standard deviations.}
\label{tb:participants_info}
\begin{tabular}{lcccccc}
    \toprule
    Scenario & Age & Extroversion & Agreeableness & Conscientiousness & Neuroticism & Openness \\
    \midrule
    Recommendation (Sec.~\ref{sec:recsys}) & 41.0 (13.5) & 0.52 (0.26) & 0.74 (0.18) & 0.76 (0.18) & 0.37 (0.21) & 0.72 (0.19) \\
    Opinion Formation (Sec.~\ref{sec:opinion}) & 42.2 (13.0) & 0.50 (0.25) & 0.71 (0.20) & 0.77 (0.19) & 0.40 (0.23) & 0.70 (0.19) \\
    Advice (Sec.~\ref{sec:advice}) & 41.3 (12.1) & 0.49 (0.25) & 0.76 (0.18) & 0.81 (0.17) & 0.37 (0.23) & 0.72 (0.19) \\
    \bottomrule
\end{tabular}
\end{table*}

\begin{table*}[t]
    \centering
    \caption{Examples of advertisements before and after tone conversion}
    \label{tb:tone_recommendation}
    \small
    \begin{tabular}{lp{12.5cm}}
        \toprule
        Tone & Explanations (Advertisements) \\
        \midrule
        Original & Prepare for an adrenaline-fueled journey in Crimson Horizon! In a dystopian future, ex-marine Zara Kane wages war against a tyrannical techno-empire. With breathtaking direction by the visionary auteur known as The Maestro and a powerful score by Sonic Pulse, this epic saga promises to ignite your spirit and heart. \\
        Neutral & Crimson Horizon follows ex-marine Zara Kane in a dystopian future as she battles a techno-empire. Directed by The Maestro and featuring a score by Sonic Pulse, this film presents a dramatic narrative. \\
        Formal & Embark on an exhilarating adventure with Crimson Horizon. Set in a dystopian future, it follows ex-marine Zara Kane's battle against a despotic techno-empire. Directed by the esteemed Maestro and accompanied by Sonic Pulse's compelling score, this epic narrative is sure to captivate and inspire. \\
        Authoritative & Brace yourself for Crimson Horizon. In a dystopian future, ex-marine Zara Kane confronts a tyrannical techno-empire. Directed by the visionary Maestro and powered by Sonic Pulse's electrifying score, this epic saga is set to ignite both spirit and heart. \\
        Humorous & Get ready to buckle up for Crimson Horizon! In a not-so-peachy future, ex-marine Zara Kane takes on a techno-empire with more sass than a cat in a laser pointer factory. Directed by The Maestro and scored by Sonic Pulse, this epic will rock your socks and your heart! \\
        Romantic & Embark on a passionate odyssey with Crimson Horizon! In a dystopian realm, brave Zara Kane battles a merciless techno-empire. Under the masterful vision of The Maestro and the soul-stirring melodies of Sonic Pulse, this enthralling saga will set your heart aflame and stir your very soul. \\
        \bottomrule
    \end{tabular}
\end{table*}

\subsection{Ethical Concerns}\label{subsec:ethical}
Before starting the user experiments, participants were informed about the experiments, its risks, and their rights.
They were asked to provide consent for the use of their demographic data and responses in a nonidentifiable form for research purposes.
The user experiments were conducted only with consenting participants.
Participants were allowed to stop the user experiments at any time.
We collected worker IDs on MTurk to identify each participant; however, this information was only used to pay the participation rewards. 
Additionally, the collected data were  used only for research purposes and were not shared with any other party to preserve the privacy of the participants.
The experimental protocol was approved by the institutional ethics committee (approval ID: 24B-08).

\section{User Experiment 1: Recommendation Scenario}\label{sec:recsys}

\subsection{Experimental Design}\label{subsec:design1}

\subsubsection{Dataset Generation}\label{subsubsec:dataset1}

Because existing movie datasets (e.g., IMDb~\cite{maas-EtAl:2011} and the Amazon review dataset~\cite{ni2019justifying}) do not include advertisements, and to avoid copyright-related concerns, we generated a fictional movie dataset using an LLM.
By using fictional movie items, we eliminate the need to check whether participants knew or had previously watched the movies, thereby eliminating bias from prior knowledge.
Based on a previous study~\cite{okoso2024impact}, we used LLMs to create 35 fictional movie items (five items for each of the seven genres), each including a title, a brief plot, and a poster image. Additionally, we generated advertisements for these items in five different tones: neutral, formal, humorous, authoritative, and romantic.
Details on how to generate each item of information and advertisements are presented in Appendix~\ref{app:dataset_recsys}.
Examples of the advertisements before and after tone conversion are summarized in Table~\ref{tb:tone_recommendation}.

\begin{figure}[t]
    \centering
    \includegraphics[width=0.98\linewidth]{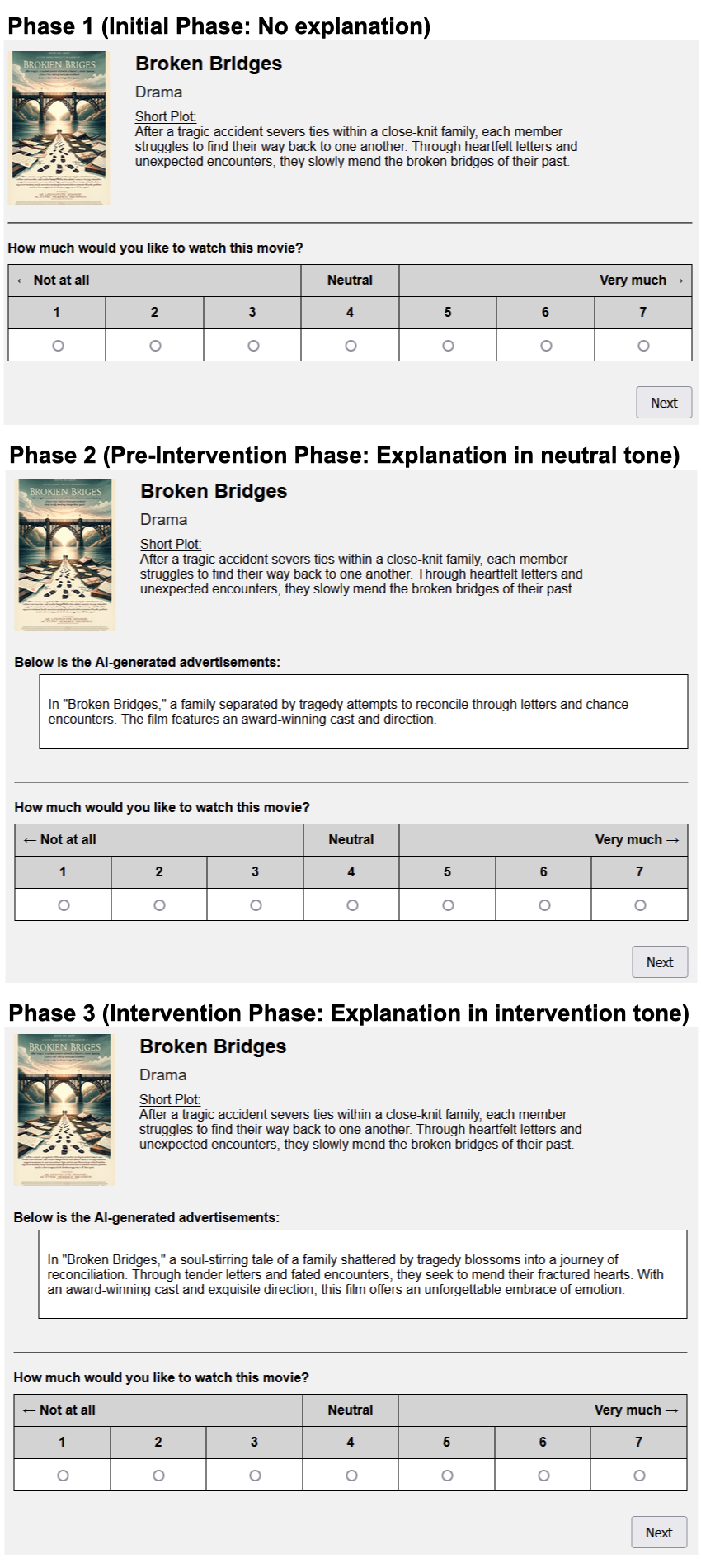}
    \caption[screenshot recommendation]{Recommendation scenario: Screen presented to users in each phase. AI-generated advertisements were provided in a neutral tone in Phase 2 and in an intervention tone in Phase 3.}
    \Description{This figure shows an example of an experimental screen for the recommendation scenario. AI-generated advertisements were displayed alongside the movie item information in Phases 2 and 3.}
    \label{fig:screenshot_recommendation}
\end{figure}

\subsubsection{Experiment Setup and Task}\label{subsubsec:setup1}

Participants were randomly assigned to one of the intervention tones (neutral, formal, humorous, authoritative, or romantic).
For each task, they were instructed to rate, on a 7-point Likert scale, how much they would like to watch the presented movie (1: Not at all, 4: Neutral, 7: Very much).
Each participant rated one movie per task, with seven movies randomly assigned to each participant from a pool of 35 movies.

Figure~\ref{fig:screenshot_recommendation} shows an example of the user experiment screen.
In Phase 1, participants rated the movie based only on movie information, which included the title, genre, short plot, and poster image.
In Phase 2, advertisements in a neutral tone were presented alongside the movie information, and in Phase 3, advertisements were presented in the assigned intervention tone.
All advertisements were clearly labeled as AI-generated.
The same seven movies were used across all phases (initial, pre-intervention, and intervention).

\subsection{Results}\label{subsec:result1}

We used the average score difference $d_u^{\rm exp}$, representing the effect of explanation presence, and $d_u^{\rm tone}$, representing the effect of explanation tone, as the evaluation metrics.
Let $s_{i, u}^{x}$ denote the score for the extent to which participant $u$ wanted to watch the movie in task $i \in I_u$ during phase $x \in \{1, 2, 3\}$, where $I_u$ is the set of tasks assigned to participant $u$.
Subsequently, $d_u^{\rm exp}$ and $d_u^{\rm tone}$ are calculated as follows:
\begin{equation}
    d_u^{\rm exp} = \frac{1}{|I_u|} \sum_{i \in I_u} (s_{i, u}^2 - s_{i, u}^1),~~~~~ d_u^{\rm tone} = \frac{1}{|I_u|} \sum_{i \in I_u} (s_{i, u}^3 - s_{i, u}^2),
    \nonumber
\end{equation}
where $|I_u|$ denotes the number of elements in $I_u$.

First, we performed an ANOVA on $d_u^{\rm exp}$ to confirm that no bias existed among intervention groups (Sec.~\ref{subsubsec:phase1to2}).
Because all groups viewed explanations in a neutral tone during Phase 2, no significant differences were expected among the groups.
Next, to examine the overall effect of tone across participants, we conducted an ANOVA on $d_u^{\rm tone}$ (Sec.~\ref{subsubsec:phase2to3}).
Additionally, we investigated how the effect of tone interacted with user attributes by plotting $d_u^{\rm tone}$ against each user attribute and calculating Pearson's correlation coefficients (Sec.~\ref{subsubsec:phase2to3_user}).
To determine the presence of significant correlations, we tested the null hypothesis that the correlation coefficients would be zero.
For gender, which is a categorical variable, we used a two-way ANOVA to analyze the differences in distribution.
All tests were conducted with a significance level of $5\%$.

\subsubsection{Bias Analysis Among Intervention Groups (Phase 1 to 2)}\label{subsubsec:phase1to2}

An ANOVA on $d_u^{\rm exp}$ revealed no significant differences among the intervention groups ($F = 0.33, p = 0.85$).
This result indicates that no bias existed among the participants assigned to the different intervention groups.

\begin{figure*}[t]
    \centering
    \includegraphics[width=0.85\linewidth]{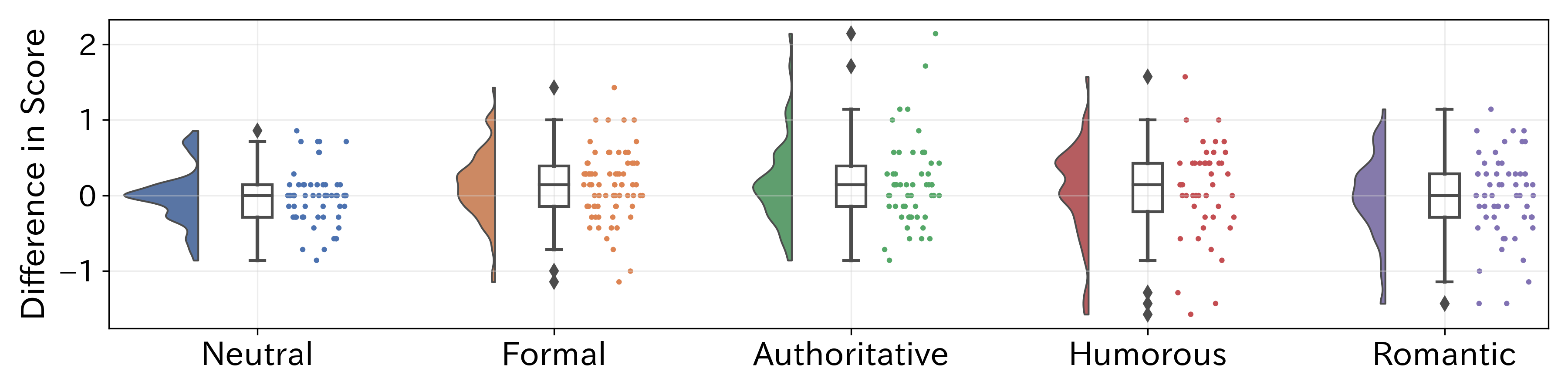}
    \caption[]{Recommendation scenario: Distributions of score differences by tone (Phases 2 to 3). The violin and box plots show $d_u^{\rm tone}$ across five tone interventions. }
    \Description{The figure shows a series of violin and box plots depicting the distribution of score differences ($d_u^{\rm tone}$) for participants across the five intervention tones: Neutral, Formal, Authoritative, Humorous, and Romantic. The horizontal axis indicates the different tone categories, whereas the vertical axis represents the score differences. Each plot includes both the distribution shape (violin) and the box plot that summarizes the interquartile range, median, and potential outliers.}
    \label{fig:movie_score_change_2to3}
\end{figure*}

\begin{figure*}[t]
    \centering
    \includegraphics[width=0.9\linewidth]{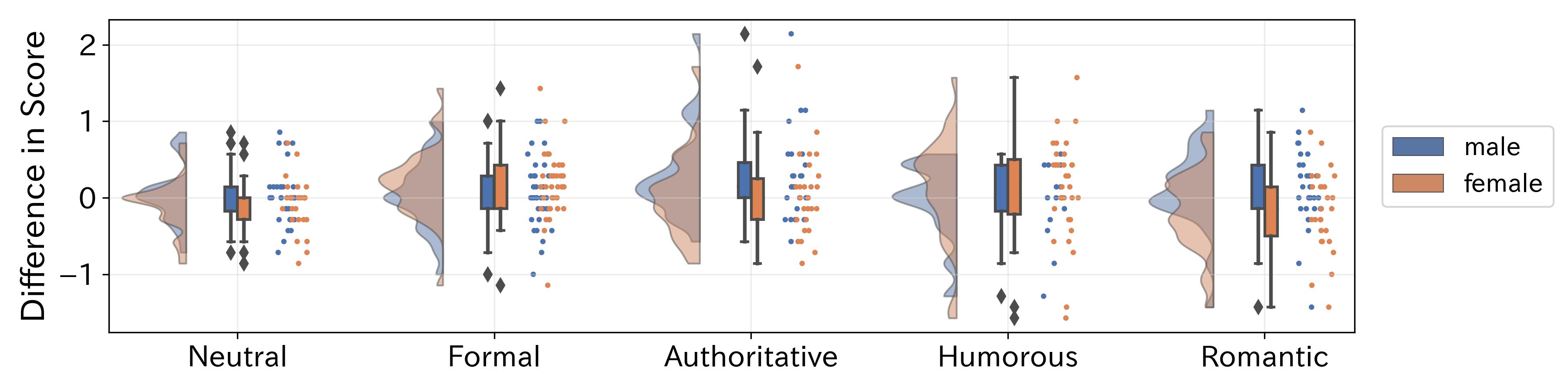}
    \caption[screenshot recommendation]{Recommendation scenario: Interaction of tone and gender on score differences (Phases 2 to 3). The plot illustrates the distribution of $d_u^{\rm tone}$ across tone interventions, divided by gender (male and female).}
    \Description{This figure presents the distribution of score differences across tone interventions, divided by gender (male and female) for the recommendation scenario. The tones include Neutral, Formal, Authoritative, Humorous, and Romantic. The horizontal axis lists the tones, and the vertical axis shows the score differences ($d_u^{\rm tone}$). Each tone group is represented with side-by-side violin and box plots for both male (blue) and female (orange) participants, showing the spread and concentration of the score differences for each gender.}
    \label{fig:movie_score_gender}
\end{figure*}

\subsubsection{Overall Effect of Tone across Participants (Phase 2 to 3)}\label{subsubsec:phase2to3}

Figure~\ref{fig:movie_score_change_2to3} shows the distribution of $d_u^{\rm tone}$ across the intervention groups.
The horizontal axis represents the intervention tones and the vertical axis represents $d_u^{\rm tone}$.
In the Formal, Authoritative, and Humorous groups, an upward trend was observed in $d_u^{\rm tone}$, indicating that participants tended to increase their scores.
We conducted an ANOVA to examine whether the effect of tone differed between intervention groups.
The results showed no significant difference ($F = 1.60, p = 0.18$).
Although no significant overall effect of tone was observed, the tone could possibly affect participants differently depending on specific user attributes.
Therefore, we analyze the interactions between tone and user attributes in the next subsection.

\subsubsection{Interaction Between Tone and User Attributes (Phases 2 to 3)}\label{subsubsec:phase2to3_user}

\paragraph{Gender}

Figure~\ref{fig:movie_score_gender} illustrates the distribution of $d_u^{\rm tone}$ by gender across the intervention tone groups. 
A two-way ANOVA for $d_u^{\rm tone}$ and gender revealed a significant interaction ($F(4, 282) = 2.44, p = 0.047$).
While the Tukey HSD post-hoc test showed no significant differences across tones, the most notable trend was observed in the Romantic group ($t = 2.31$, adjusted $p = 0.12$).
As shown in Fig.~\ref{fig:movie_score_gender}, male participants in the Romantic group tended to increase their scores, whereas female participants tended to decrease their scores.

\paragraph{Age}

\begin{figure*}[t]
    \centering
    \includegraphics[width=1.0\linewidth]{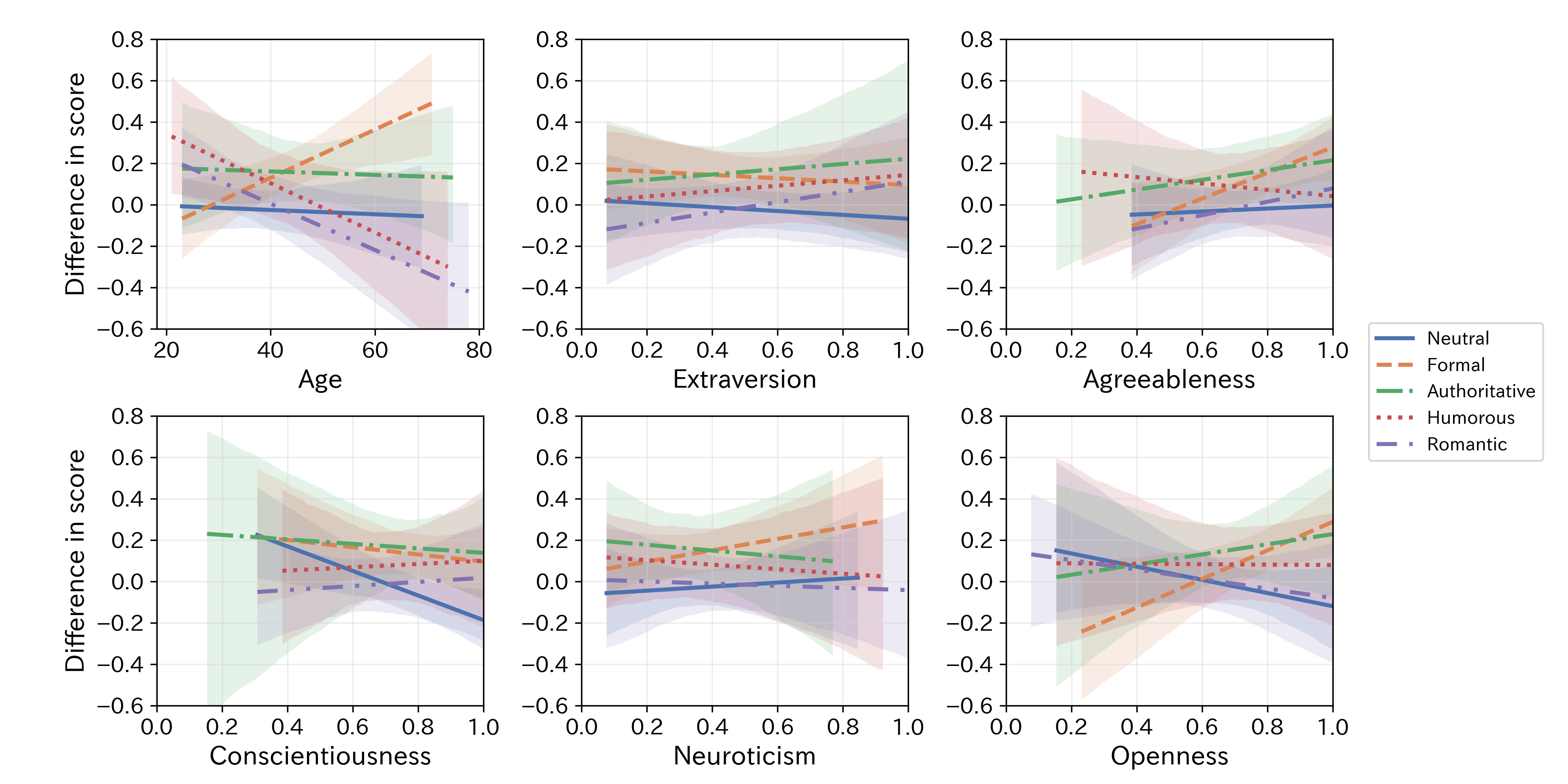}
    \caption[screenshot recommendation]{Recommendation scenario: Correlation between age, personality traits, and score differences by tone (Phase 2 to 3). The top-left graph shows the relationship between age and $d_u^{\rm tone}$, whereas the remaining graphs show the relationship for personality traits.}
    \Description{The figure contains six line plots, with the top-left graph showing the relationship between age and score differences, and the remaining five graphs illustrating the relationships between score differences and various personality traits: Extroversion, Agreeableness, Conscientiousness, Neuroticism, and Openness. The horizontal axis in each plot represents either age or personality trait levels (ranging from 0 to 1 for personality traits and from 20 to 80 for age), whereas the vertical axis represents the difference in scores.}
    \label{fig:movie_score_correlation_attributes}
\end{figure*}

Figure~\ref{fig:movie_score_correlation_attributes} illustrates the relationship between age and $d_u^{\rm tone}$.
Correlation tests revealed significant relationships in the Formal and Romantic groups.
Specifically, a positive correlation was observed in the Formal group ($r = 0.31$, adjusted $p = 0.033$), whereas a negative correlation was observed in the Romantic group ($r = -0.31$, adjusted $p = 0.046$).
Additionally, a marginally significant trend was observed in the Humorous group ($r = -0.28$, adjusted $p = 0.09$).
In summary, older participants tended to decrease their scores for the romantic and humorous tones, while increasing their scores for the formal tone.

\paragraph{Personality}

Figure~\ref{fig:movie_score_correlation_attributes} illustrates the relationship between each personality trait and $d_u^{\rm tone}$.
While correlation tests revealed no significant relationships in any tone group or personality trait, some trends were observed.
In the Formal group, agreeableness ($r = 0.26$, adjusted $p = 0.12$) and openness ($r = 0.27$, adjusted $p = 0.09$) showed near-significant trends, with higher levels of these traits associated with increased scores.
Similarly, in the Neutral group, conscientiousness exhibited a near-significant trend ($r = -0.29$, adjusted $p = 0.12$), where higher levels of conscientiousness were associated with decreased scores.

\subsubsection{Qualitative Analysis}\label{subsubsec:discussions1}

To understand how users perceived the AI-generated explanations and their impact on decision-making, we analyzed the responses to open-ended questions Q1 and Q2 from the experiment (See Table~\ref{tb:questionnaire}).

\paragraph{Impact of Explanations on Decision-making (Q1)}
A total of 110 participants ($37\%$) expressed either positive or negative opinions about the explanations but ultimately stated that it did not influence their decision.
Most felt that the explanations did not provide sufficient information relevant to their needs.
Some participants mentioned that they chose movies based solely on genre and did not consider the explanation, whereas others remarked that it felt unnatural for an AI, which supposedly lacks emotions, to promote a movie.
This suggests that the explanations might not have engaged users sufficiently, potentially obscuring the influence of tone.

\paragraph{Impact of Tone Changes on Decision-making (Q2)}
It was evident that even when the same tone was assigned, participants responded differently.
For instance, one participant (P131) who was shown the Romantic tone stated, ``In Phase 3, I selected a higher score because richer and more expressive language was used.''
However, another participant (P115) said, ``The language in Phase 3 felt overdone, which made me less interested in the movie.''
This demonstrates that, even with the same tone, the perceptions of users and the effect on their decision-making varied from person to person.

\begin{table*}[t]
    \centering
    \caption{Examples of opinions before and after tone conversion}
    \label{tb:tone_opinion}
    \small
    \begin{tabular}{lp{12.5cm}}
        \toprule
        Tone & Explanations (Opinions) \\
        \midrule
        Original & Eliminating student loan debt through forgiveness or bankruptcy can significantly reduce financial stress for graduates, enabling them to focus on their careers and contribute more effectively to the economy. \\
        Neutral & Eliminating student loan debt through forgiveness or bankruptcy can reduce financial stress for graduates, allowing them to concentrate on their careers and contribute more effectively to the economy.\\
        Formal & The elimination of student loan debt through forgiveness or bankruptcy has the potential to substantially alleviate financial stress for graduates, thereby allowing them to concentrate more fully on their careers and contribute more effectively to the economy.\\
        Authoritative & Eliminating student loan debt through forgiveness or bankruptcy will decisively reduce financial stress for graduates, empowering them to focus on their careers and enhancing their contributions to the economy.\\
        Casual & Wiping out student loans through forgiveness or bankruptcy can really ease the financial stress for grads, letting them dive into their careers and boost the economy big time.\\
        Humorous & Wiping out student loan debt is like giving graduates a financial spa day\u2014they can finally focus on their careers without that debt monster breathing down their necks, and hey, the economy might actually start to look like it's had its morning coffee!\\
        \bottomrule
    \end{tabular}
\end{table*}

\section{User Experiment 2: Opinion Formation Scenario}\label{sec:opinion}

\subsection{Experimental Design}\label{subsec:design2}

\subsubsection{Dataset Generation}\label{subsubsec:dataset2}

We manually selected 11 topics from the ProCon website~\footnote{https://www.procon.org/}, ten of which were used for user experiments and one for the attention check.
The topics were categorized into education, healthcare, environment \& animals, digital life, and sports.
Examples include \textit{Should Students Have to Wear School Uniforms?}, \textit{Should Employers Be Able to Mandate Vaccinations?}, and \textit{Should Zoos Exist?}

Opinions on each topic were generated to provide explanations to the participants.
To observe changes in the participants' behavior from their initial intentions, we only used opinions that clearly favored the topic. 
The detailed generation procedure is described in Appendix~\ref{app:dataset_opinion}.
Examples of the generated opinions before and after tone conversion are summarized in Table~\ref{tb:tone_opinion}.

\begin{figure*}[t]
    \centering
    \includegraphics[width=0.7\linewidth]{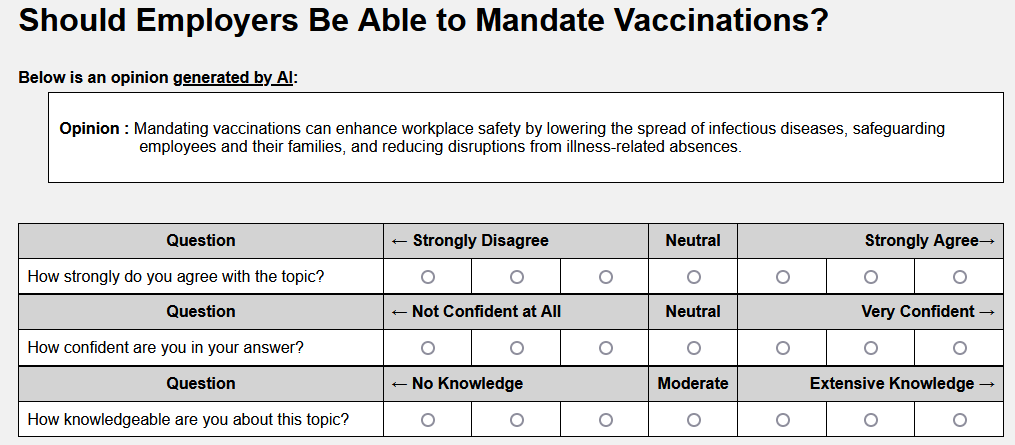}
    \caption[screenshot opinion]{Opinion Formation Scenario: Screen presented to users in the pre-intervention phase. An AI-generated supportive opinion was presented in a neutral tone.}
    \label{fig:screenshot_opinion}
    \Description{This figure shows an example of an experimental screen for the pre-intervention phase of the opinion formation scenario. An AI-generated supportive opinion was presented in a neutral tone.}
\end{figure*}

\subsubsection{Experiment Setup and Task}\label{subsubsec:setup2}

Participants were randomly assigned to one of the intervention tones (neutral, formal, humorous, authoritative, or casual).
For each task, the participants rated their stance on a presented topic using a 7-point Likert scale (1: Strongly Disagree, 4: Neutral, 7: Strongly Agree).
They also rated their confidence in their stance on a 7-point Likert scale (1: Not Confident at All, 4: Neutral, 7: Very Confident) and their level of knowledge about the topic on a 7-point Likert scale (1: Not Knowledgeable at All, 4: Moderate, 7: Very Knowledgeable). 
Each participant rated one topic per task, with five topics randomly assigned from a pool of ten topics.

Figure~\ref{fig:screenshot_opinion} shows an example of the user experiment screen.
In the initial phase (Phase 1), participants received no explanations.
In the pre-intervention phase (Phase 2), supportive opinions were presented in a neutral tone, and in the intervention phase (Phase 3), supportive opinions were presented in the assigned intervention tone.
All presented opinions were labeled as AI-generated.
The participants rated the same five topics across all phases (initial, pre-intervention, and intervention).

\subsection{Results}\label{subsec:result2}

We used the differences in average agreement score ($d_u^{\rm exp, stance}$, $d_u^{\rm tone, stance}$), average confidence score ($d_u^{\rm exp, conf}$, $d_u^{\rm tone, conf}$), and average knowledge score ($d_u^{\rm exp, know}$, $d_u^{\rm tone, know}$) as evaluation metrics.
Here, $d_u^{\rm exp, \cdot}$ represents the effect of explanation presence, and $d_u^{\rm tone, \cdot}$  represents the effect of the explanation tone.
The calculation method for these metrics and the analysis procedure are identical to those used in the recommendation scenario (see Sec.~\ref{subsec:result1}).

\subsubsection{Bias Analysis Among Intervention Groups (Phase 1 to 2)}\label{subsubsec:phase1to2_2}

ANOVAs on $d_u^{\rm exp, stance}$, $d_u^{\rm exp, conf}$, and $d_u^{\rm exp, know}$ revealed no significant differences among the intervention groups ($F = 0.33$, $p = 0.85$ for $d_u^{\rm exp, stance}$; $F = 0.27$, $p = 0.90$ for $d_u^{\rm exp, conf}$; and $F = 1.08$, $p = 0.37$ for $d_u^{\rm exp, know}$).
These results confirm that no biases existed among participants assigned to different intervention groups.

\begin{figure*}[tb]
    \centering
    \begin{minipage}[tb]{\linewidth}
        \centering
        \includegraphics[width=0.85\linewidth]{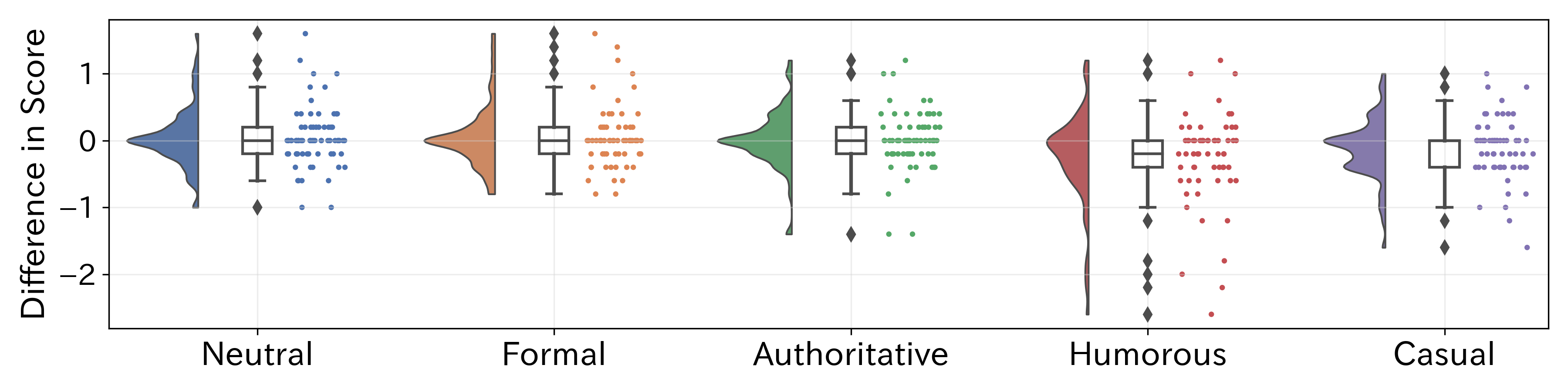}
        \subcaption[]{Distribution of the agreement score differences by tone ($d_u^{\rm tone, stance}$).}
        \label{fig:opinion_agreement_change_2to3}
    \end{minipage}
    \vfill
    \begin{minipage}[tb]{\linewidth}
        \centering
        \includegraphics[width=0.85\linewidth]{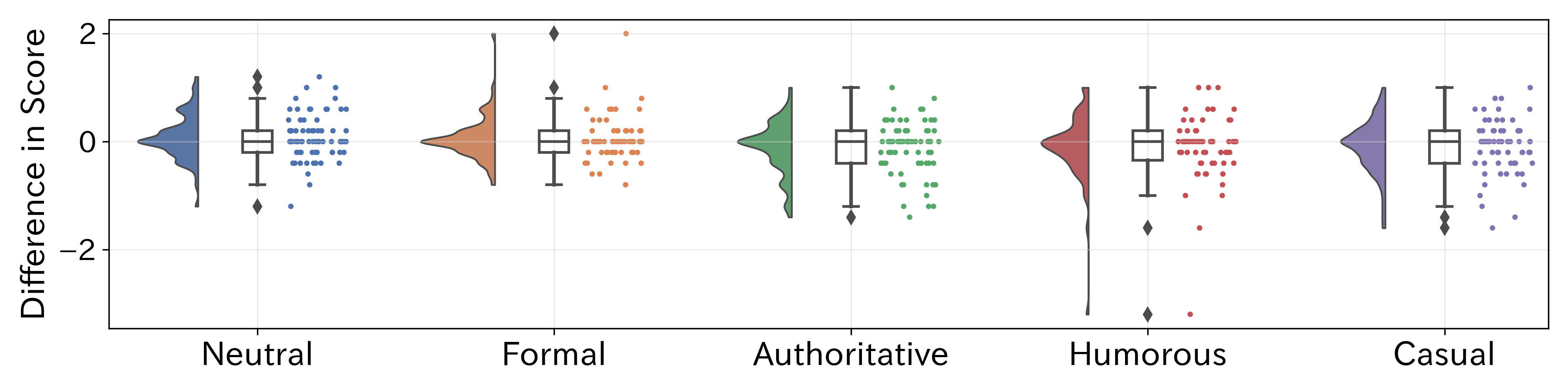}
        \subcaption[]{Distribution of the confidence score differences by tone ($d_u^{\rm tone, conf}$).}
        \label{fig:opinion_confidence_change_2to3}
    \end{minipage}
    \vfill
    \begin{minipage}[tb]{\linewidth}
        \centering
        \includegraphics[width=0.85\linewidth]{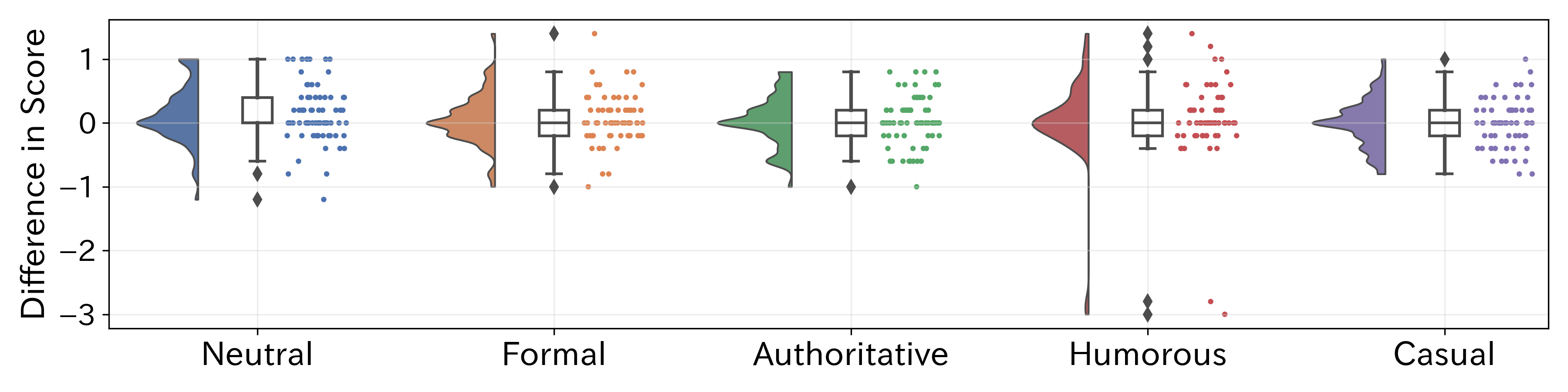}
        \subcaption[]{Distribution of the knowledge score differences by tone ($d_u^{\rm tone, know}$).}
        \label{fig:opinion_knowledge_change_2to3}
    \end{minipage}
    \caption{Opinion formation scenario: Score differences by tone for agreement, confidence, and knowledge (Phases 2 to 3). The violin and box plots show (a) $d_u^{\rm tone, stance}$, (b) $d_u^{\rm tone, conf}$, and (c) $d_u^{\rm tone, know}$ across five tone interventions.}
    \Description{The figure contains three subfigures, each representing the distribution of score differences across the five tone interventions: Neutral, Formal, Authoritative, Humorous, and Casual. Each subfigure uses violin and box plots to depict the range and concentration of score differences for a specific metric.}
\end{figure*}

\subsubsection{Overall Effect of Tone across Participants (Phases 2 to 3)}\label{subsubsec:phase2to3_2}

Figures~\ref{fig:opinion_agreement_change_2to3}, \ref{fig:opinion_confidence_change_2to3}, and~\ref{fig:opinion_knowledge_change_2to3} illustrate the distributions of $d_u^{\rm tone, stance}$, $d_u^{\rm tone, conf}$, and $d_u^{\rm tone, know}$, respectively, across the intervention groups.
ANOVAs revealed significant differences for $d_u^{\rm tone, stance}$ ($F = 4.40, p = 0.002$) and $d_u^{\rm tone, conf}$ ($F = 2.45$, $p = 0.046$).
However, no significant difference was observed for $d_u^{\rm tone, know}$ ($F = 0.89$, $p = 0.47$).
These results suggest that tone influenced agreement and confidence scores but had no noticeable effect on perceived knowledge.

The Tukey HSD post-hoc test was conducted for both $d_u^{\rm tone, stance}$ and $d_u^{\rm tone, conf}$.
For $d_u^{\rm tone, stance}$, the results showed significant differences between the Authoritative and Humorous ($p = 0.020$), Formal and Humorous ($p = 0.007$), as well as Humorous and Neutral ($p = 0.006$) groups.
For $d_u^{\rm tone, conf}$, no significant difference was found between any pairs, although a near-significant trend was observed between the Formal and Humorous groups ($p = 0.08$).

\begin{figure*}[t]
    \centering
    \includegraphics[width=1.0\linewidth]{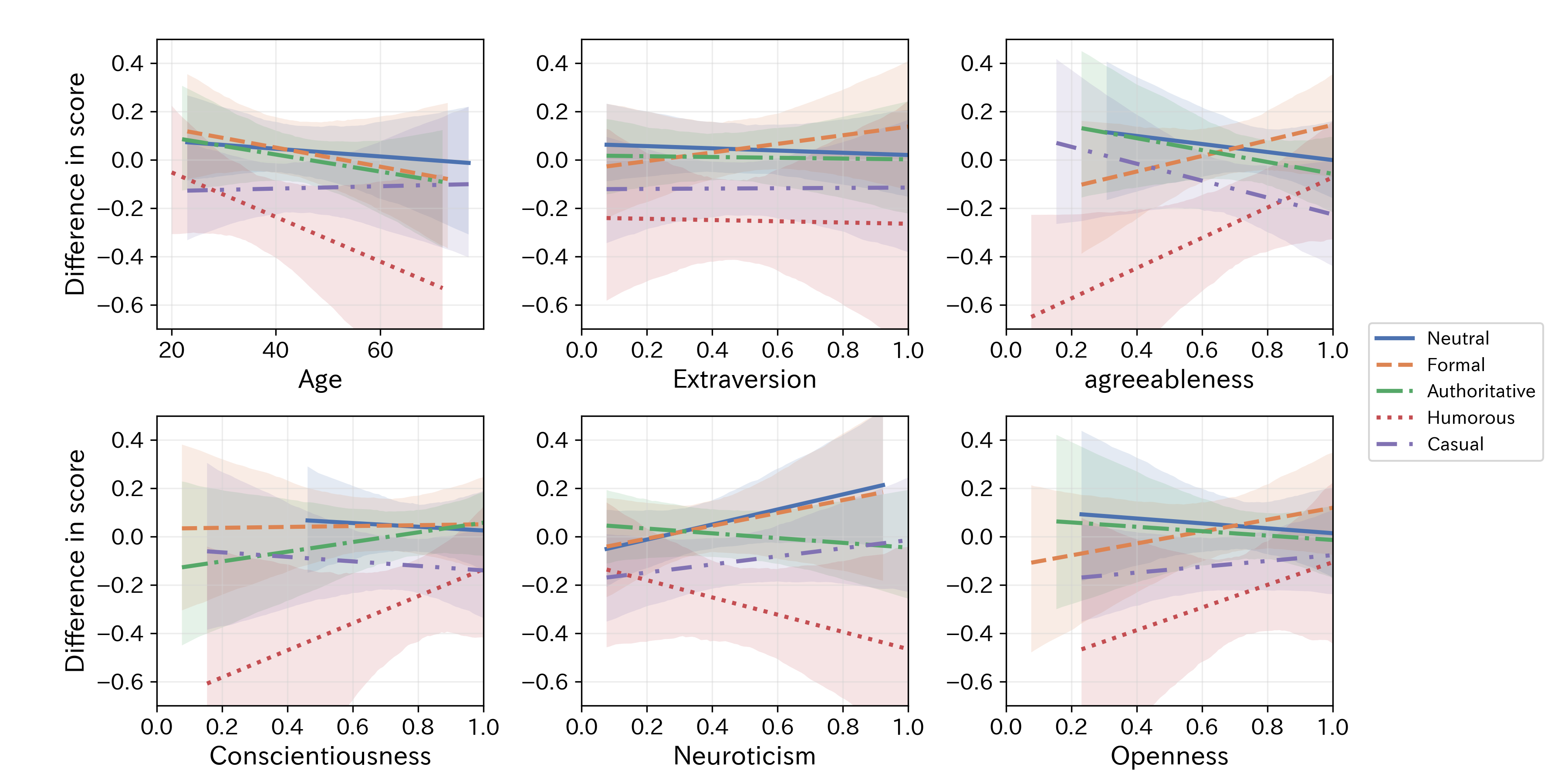}
    \caption{Opinion formation scenario: Correlation between age, personality traits, and $d_u^{\rm tone, stance}$. The top-left graph shows the relationship between age and $d_u^{\rm tone, stance}$, whereas the remaining graphs illustrate the relationship for five personality traits.}
    \Description{The figure consists of six line plots, where the top-left graph displays the relationship between age and the agreement score differences ($d_u^{\rm tone, stance}$), and the remaining five graphs show the relationships between score differences and the following personality traits: Extroversion, Agreeableness, Conscientiousness, Neuroticism, and Openness.}
    \label{fig:opinion_agreement_correlation_attributes}
\end{figure*}

\subsubsection{Interaction Between Tone and User Attributes (Phases 2 to 3)}\label{subsubsec:phase2to3_user_2}

\paragraph{Gender}

Two-way ANOVAs revealed no significant relationship between the tone and gender for any of the metrics: $F = 0.27, p = 0.90$ for $d_u^{\rm tone, stance}$, $F = 0.27, p = 0.90$ for $d_u^{\rm tone, conf}$, and $F = 0.27, p = 0.90$ for $d_u^{\rm tone, know}$.
These results indicate that the interaction between tone and gender did not have a significant effect on any of the scores.

\paragraph{Age}

Figure~\ref{fig:opinion_agreement_correlation_attributes} illustrates the relationship between age and $d_u^{\rm tone, stance}$.
Correlation tests revealed no significant correlations among any of the metrics or tone groups.
However, positive correlations were observed for $d_u^{\rm tone, conf}$ ($r = 0.24$, adjusted $p = 0.19$) and for $d_u^{\rm tone, know}$ ($r = 0.24$, adjusted $p = 0.21$) in the Authoritative group.
These results suggest that confidence and perceived knowledge increased with age for the authoritative tone.

\paragraph{Personality}

Figure~\ref{fig:opinion_agreement_correlation_attributes} illustrates the relationship between each personality trait and $d_u^{\rm tone, stance}$.
Correlation tests revealed no significant relationships for any of the metrics or personality traits.
However, some trends were observed in certain tone groups for $d_u^{\rm tone, conf}$; specifically, positive correlations were found for openness in the Authoritative ($r = 0.21$, adjusted $p = 0.21$) and Casual ($r = 0.22$, adjusted $p = 0.21$) groups.

\subsubsection{Qualitative Analysis}\label{subsubsec:discussions２}

\paragraph{Impact of Explanations on Decision-making (Q1)}
A total of 52\% of participants responded that it influenced or informed their decision.
Conversely, 36\% of the participants reported that it did not influence or inform their decision.
Those who felt influenced by the explanations often mentioned that they provided new perspectives, helped them think more deeply about their decisions, and increased their confidence.
In contrast, participants who were not influenced frequently stated that the information was already familiar to them or that they made their decisions based on their own beliefs despite reading the explanation.
This suggests that tone may have slightly impacted the participants who were not influenced by the AI-generated explanation.

\paragraph{Impact of Tone Changes on Decision-making (Q2)}
Excluding participants who did not notice the tone change or provided irrelevant answers, 179 participants remained.
Among them, 88 said it did not affect their decisions while 74 said it did. 
Among those who were influenced, the participants assigned to the Humorous tone, such as P56, P57, and P59, mentioned that the tone made the explanation more engaging and increased their interest in the topic.
However, P115 said that the tone was too enthusiastic, making them more cautious about the explanation.
For the Formal tone, P107 felt that the explanation was more detailed and increased their confidence, whereas P267 said that the wording was vague, making them more skeptical of the explanation.
In the Authoritative group, P83 reported that the explanation seemed knowledgeable and boosted their confidence, whereas P31 found the explanation arrogant and unappealing.
These responses show that even with the same tone, participants perceived and reacted to the explanations differently.

\section{User Experiment 3: Advice Scenario}\label{sec:advice}

\subsection{Experimental Design}\label{subsec:design3}

\subsubsection{Dataset Generation}\label{subsubsec:dataset3}

We used the COMPAS dataset~\cite{compas} and generated explanations using an LLM.
The COMPAS dataset includes profile information and recidivism risk scores for over 18,000 criminal defendants in Florida, which were used to score their likelihood of reoffending within two years.
The recidivism risk scores range from 1 to 10, with 10 indicating the highest risk.
Additionally, the COMPAS dataset categorizes these scores into three levels: low (1--4), medium (5--7), and high (8--10) risks.

In this task, we used the following profile information based on previous work~\cite{dressel2018accuracy}: age, gender, charge, charge degree (felony or misdemeanor), number of prior adult offenses, number of juvenile felony offenses, and number of juvenile misdemeanor offenses.
From the COMPAS dataset, we randomly selected data from 100 defendants to ensure that all the profile fields were complete.
The explanations used in this task consisted of the predicted recidivism risk scores and reasoning assumed to be provided by an AI system.
We used data from 30 out of the defendants for whom the AI system's prediction matched the risk scores of the COMPAS dataset.
The detailed generation procedure is described in Appendix~\ref{app:dataset_advice}.
Examples of the generated explanations before and after tone conversion are listed in Table~\ref{tb:tone_advice}.

\begin{table*}[t]
    \centering
    \caption{Examples of advice before and after tone conversion}
    \label{tb:tone_advice}
    \small
    \begin{tabular}{lp{12.5cm}}
        \toprule
        Tone & Explanations (Advice) \\
        \midrule
        Original & The individual has a relatively low risk of recidivism due to being older (age 51) and having a minimal criminal history (only one prior non-juvenile offense). The nature of the crime (DUI) does elevate the risk slightly but not significantly.\\
        Neutral & The individual has a low risk of recidivism due to being 51 years old and having a minimal criminal history with only one prior non-juvenile offense. The nature of the crime (DUI) slightly elevates the risk but not significantly.\\
        Formal & The individual presents a relatively low risk of recidivism, attributable to their age (51) and limited criminal history (only one prior non-juvenile offense). While the nature of the crime (DUI) does moderately elevate the risk, it remains not significantly concerning.\\
        Authoritative & This individual presents a low recidivism risk, attributed to their age (51) and minimal criminal history (one prior non-juvenile offense). Although the DUI offense marginally elevates the risk, it remains insignificant.\\
        Casual & This person isn't likely to re-offend much, being 51 and with just one past non-juvenile crime. While a DUI does bump up the risk a bit, it’s not a huge deal.\\
        Humorous & At 51, this person’s more into knitting than committing crimes, with just one prior non-juvenile hiccup. Sure, the DUI adds a sprinkle of risk, but let's face it, they're more likely to be found napping than reoffending.\\
        \bottomrule
    \end{tabular}
\end{table*}

\subsubsection{Experiment Setup and Task}\label{subsubsec:setup3}

Participants were randomly assigned to one of the intervention tones (neutral, formal, humorous, authoritative, or casual).
For each task, they assessed the risk of reoffending within two years for a given defendant profile using a 10-point scale (1: Low Risk, 10: High Risk).
They also rated their confidence in their response on a 7-point Likert scale (1: Not Confident at All, 4: Neutral, 7: Very Confident).
Each participant assessed one defendant per task, with ten defendants randomly selected from a pool of 30.

\begin{figure*}[t]
    \centering
    \includegraphics[width=0.7\linewidth]{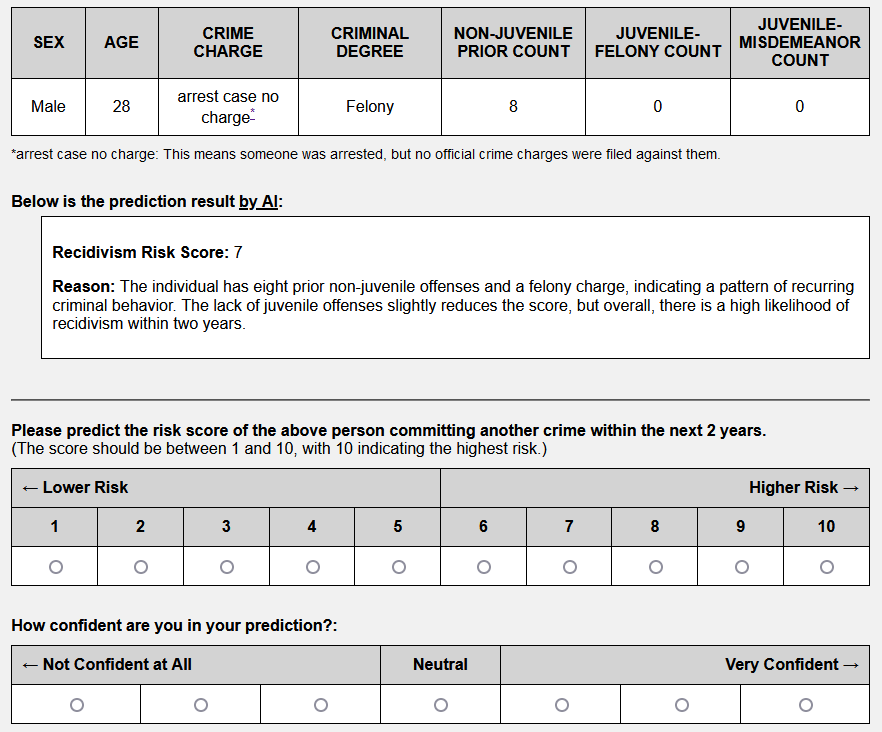}
    \caption[screenshot advice]{Advice scenario: Screen presented to users in the pre-intervention phase. Defendant profiles were shown with an AI-generated prediction of recidivism risk and an explanation in a neutral tone.}
    \label{fig:screenshot_advice}
    \Description{This figure shows an example of an experimental screen for the pre-intervention phase of the advice scenario. Defendant profiles were displayed with an AI-generated prediction of recidivism risk and an explanation in a neutral tone.}
\end{figure*}

Figure~\ref{fig:screenshot_advice} shows an example of the user experiment screen.
In Phase 1, participants were provided with only the defendant's profile, including age, gender, charges, charge severity (felony or misdemeanor), number of prior adult convictions, number of prior juvenile felony convictions, and number of prior juvenile misdemeanor convictions.
A brief description of the charges was also presented.
In Phase 2, an AI-generated predictive risk score and an explanation for the prediction were provided in a neutral tone.
In Phase 3, the AI-generated predictive risk score and explanation were presented in the assigned intervention tone.
Participants rated the same ten defendants across the three phases.

\begin{figure*}[t]
    \centering
    \begin{minipage}[tb]{\linewidth}
        \centering
        \includegraphics[width=0.85\linewidth]{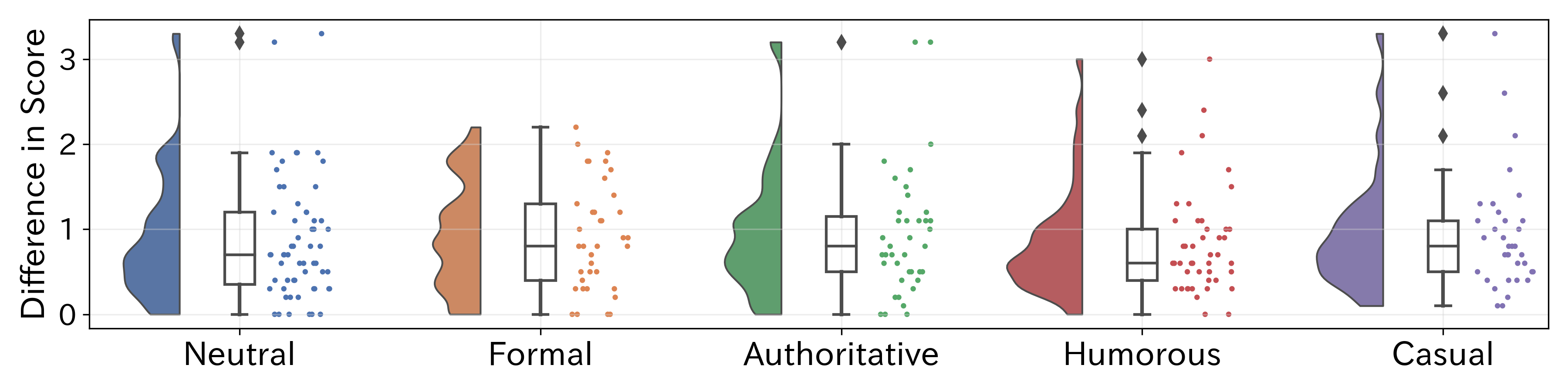}
        \subcaption[screenshot recommendation]{Distribution of the prediction risk score differences by tone ($d_u^{\rm tone, risk}$).}
        \Description{Distribution of the prediction risk score differences by tone.}
        \label{fig:advice_pred_score_abs_2to3}
    \end{minipage}
    \vfill
    \begin{minipage}[tb]{\linewidth}
        \centering
        \includegraphics[width=0.85\linewidth]{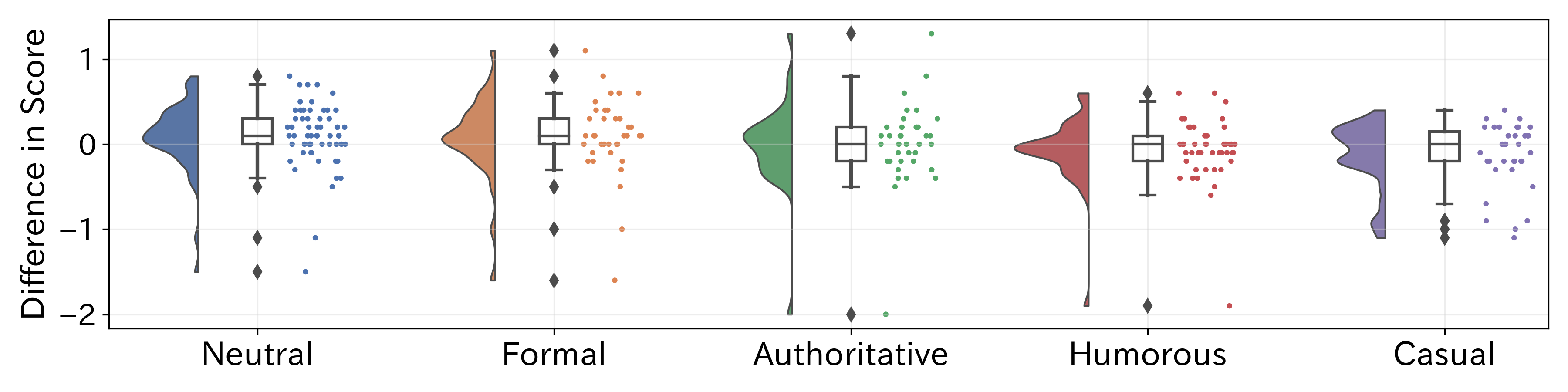}
        \subcaption[screenshot recommendation]{Distribution of the confidence score differences by tone ($d_u^{\rm tone, conf}$).}
        \Description{Distribution of the confidence score differences by tone.}
        \label{fig:advice_confidence_2to3}
    \end{minipage}
    \caption{Advice scenario: Score differences by tone for prediction risk and confidence (Phases 2 to 3). The violin and box plots show the score differences across five tone interventions for (a) $d_u^{\rm tone, risk}$ and (b) $d_u^{\rm tone, conf}$.}
    \Description{The figure consists of two subfigures comparing the distribution of prediction score differences ($d_u^{\rm tone, risk}$) and confidence score differences ($d_u^{\rm tone, conf}$) across five intervention tone groups: Neutral, Formal, Authoritative, Humorous, and Casual.}
\end{figure*}

\subsection{Results}\label{subsec:result3}

We used the average prediction score difference ($d_u^{\rm exp, risk}$, $d_u^{\rm tone, risk}$) and the average confidence score difference ($d_u^{\rm exp, conf}$, $d_u^{\rm tone, conf}$) as the evaluation metrics.
Here, $d_u^{\rm exp, \cdot}$ represents the effect of the explanation presence, and $d_u^{\rm tone, \cdot}$ represents the effect of the explanation tone.
Because the explanations included an AI-generated prediction score, $d_u^{\rm exp, risk}$ and $d_u^{\rm tone, risk}$ were calculated using absolute values to measure the extent to which the explanation text influenced the participants' own predictions, regardless of the provided score.
The calculation method for these metrics and the analysis procedure were identical to those used in the recommendation scenario (see Sec.~\ref{subsec:result1}).
Note that the larger the absolute difference in scores between Phases 2 and 3, the greater the influence of tone.

\subsubsection{Bias Analysis Among Intervention Groups (Phase 1 to 2)}\label{subsubsec:phase1to2_3}

ANOVAs on $d_u^{\rm exp, risk}$ and $d_u^{\rm exp, conf}$ revealed no significant differences among intervention groups ($F = 0.89, p = 0.47$ for $d_u^{\rm exp, risk}$; and $F = 1.63, p = 0.17$ for $d_u^{\rm exp, conf}$).
These results confirm that no biases existed among participants assigned to different intervention groups.

\subsubsection{Overall Effect of Tone Across Participants (Phases 2 to 3)}\label{subsubsec:phase2to3_3}

Figures~\ref{fig:advice_pred_score_abs_2to3} and \ref{fig:advice_confidence_2to3} illustrate the distributions of $d_u^{\rm exp, risk}$ and $d_u^{\rm exp, conf}$ across the intervention tone groups.
While an ANOVA on $d_u^{\rm exp, risk}$ showed no significant difference ($F = 0.17$, $p = 0.95$), the tone had a significant effect on $d_u^{\rm exp, conf}$ ($F = 2.52$, $p = 0.042$).

Additionally, a Tukey HSD post-hoc test was performed on $d_u^{\rm exp, conf}$; however no significant differences were found between any of the pairs.
However, a near-significant difference was observed between the Casual and Neutral groups ($p = 0.072$), suggesting that the tone may influence confidence scores in certain cases, although no overall significant pairwise differences were observed.

\subsubsection{Interaction Between Tone and User Attributes (Phase 2 to 3)}\label{subsubsec:phase2to3_user_3}

\paragraph{Gender}

\begin{figure*}[t]
    \centering
    \includegraphics[width=0.9\linewidth]{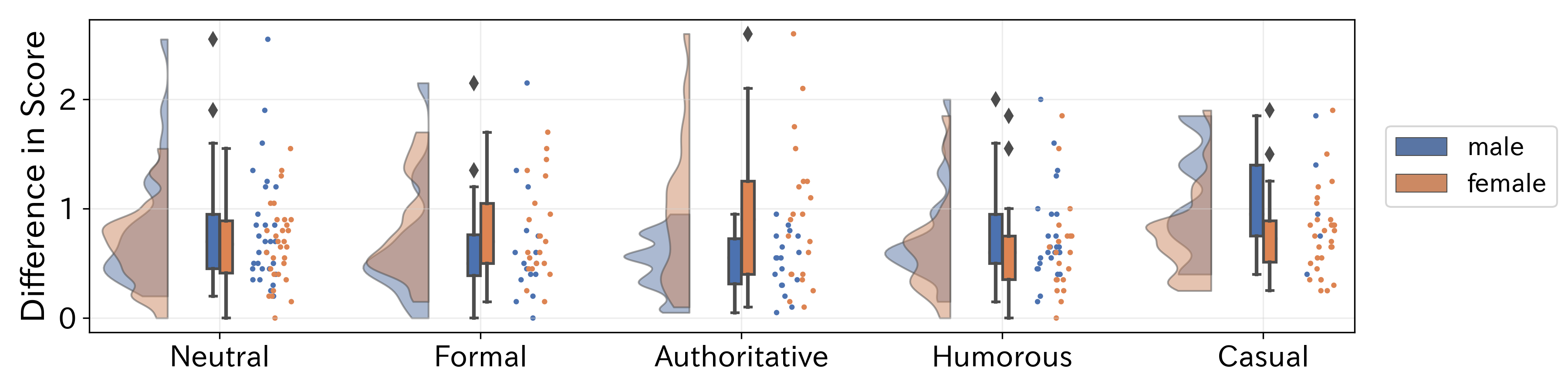}
    \caption[]{Advice scenario: Interaction of tone and gender on the prediction risk score differences (Phases 2 to 3). The plot illustrates the distribution of $d_u^{\rm tone, risk}$ across tone interventions, divided by gender (male and female).} 
    \label{fig:advice_pred_score_gender}
    \Description{This figure shows the distribution of prediction score differences ($d_u^{\rm tone, risk}$) across five tone interventions: Neutral, Formal, Authoritative, Humorous, and Casual. The data is divided by gender (male and female), with separate violin and box plots representing each gender for each tone group.}
\end{figure*}

Figure~\ref{fig:advice_pred_score_gender} illustrates the distribution of $d_u^{\rm tone, risk}$ by gender across the intervention tone groups.
A two-way ANOVA revealed a significant interaction for $d_u^{\rm tone, risk}$ ($F = 3.27$, $p = 0.013$).
Tukey HSD post-hoc tests showed a near-significant difference between males and females in the Authoritative group ($p = 0.067$). 
As shown in Fig.~\ref{fig:advice_pred_score_gender}, females were more influenced by the authoritative tone, resulting in larger prediction score changes.
A two-way ANOVA on $d_u^{\rm tone, conf}$ also revealed a significant interaction between tone and gender ($F = 3.17$, $p = 0.015$).
Although no significant differences were found in any intervention groups in the post-hoc tests, a near-significant trend was observed in the Formal group ($p = 0.051$).

\begin{figure*}[t]
    \centering
    \includegraphics[width=1.0\linewidth]{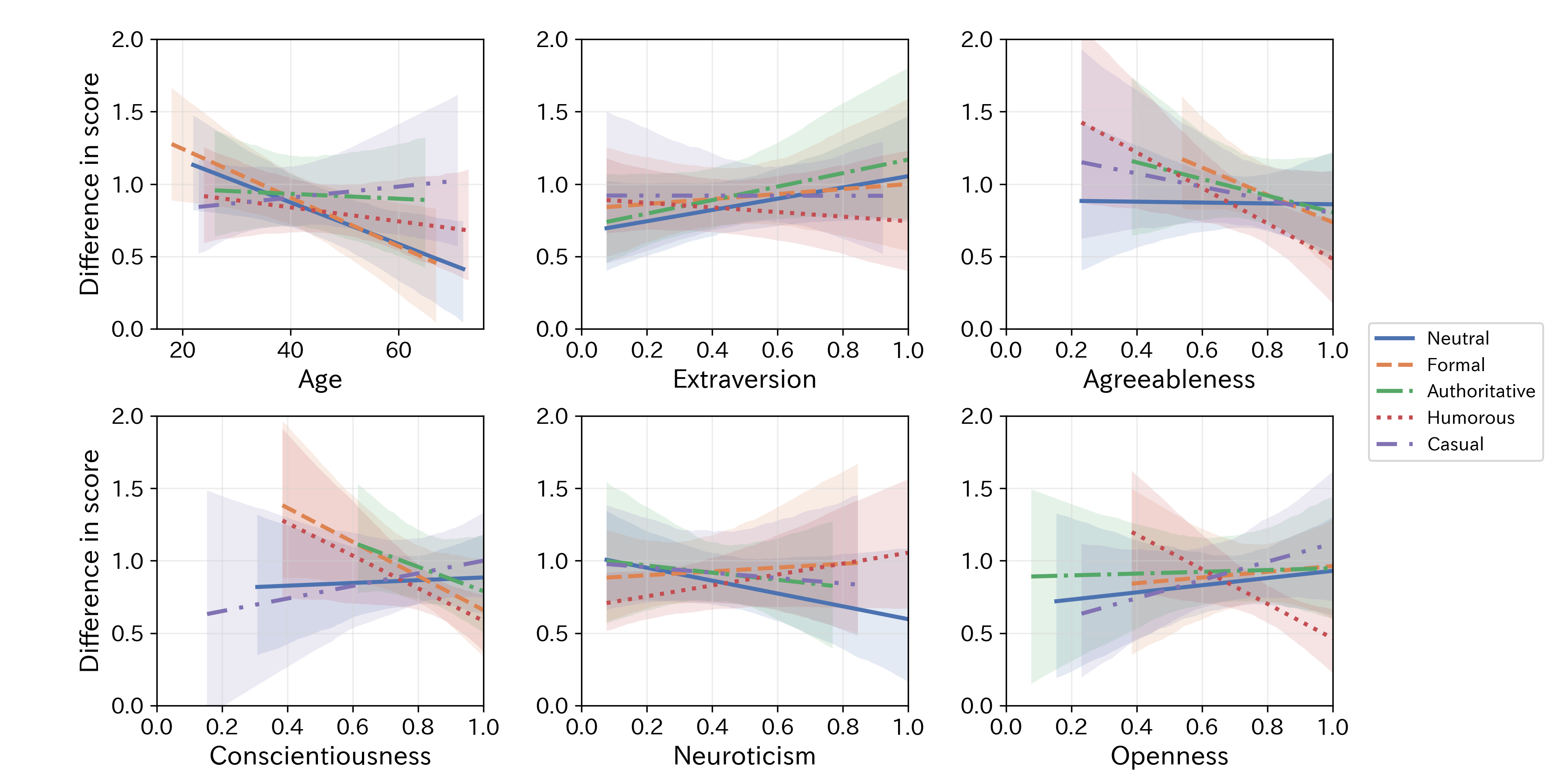}
    \caption[]{Advice scenario: Correlation between age, personality traits, and the prediction risk score differences by tone (Phase 2 to 3). The top-left graph shows the relationship between age and $d_u^{\rm tone, risk}$, whereas the remaining graphs illustrate the relationship for five personality traits.} 
    \Description{This figure contains multiple line plots, where the top-left graph shows the relationship between age and prediction score differences, whereas the remaining graphs illustrate the relationships between personality traits (Extroversion, Agreeableness, Conscientiousness, Neuroticism, and Openness) and prediction score differences.}
    \label{fig:advice_pred_score_correlation_attributes}
\end{figure*}

\paragraph{Age}

Figure~\ref{fig:advice_pred_score_correlation_attributes} illustrates the relationship between age and $d_u^{\rm tone, risk}$.
A correlation test on $d_u^{\rm tone, risk}$ revealed no significant relationships between age and tone.
However, near-significant negative correlations were observed in the Formal ($r = -0.32$, adjusted $p = 0.16$) and Neutral ($r = -0.24$, adjusted $p = 0.16$) groups.
Similarly, although a correlation test on $d_u^{\rm tone, conf}$ revealed no significant relationship, a positive correlation was observed in the Authoritative group ($r = 0.18$, $p = 0.26$).

\paragraph{Personality}

Figure~\ref{fig:advice_pred_score_correlation_attributes} illustrates the relationship between personality traits and $d_u^{\rm tone, risk}$.
Although the correlation tests on $d_u^{\rm tone, risk}$ revealed no significant relationships, some near-significant trends were observed in the Humorous group, including negative correlations with agreeableness ($r = -0.35$, adjusted $p = 0.088$), openness ($r = -0.35$, adjusted $p = 0.10$), and conscientiousness ($r = -0.35$, adjusted $p = 0.098$).
Similarly, although no statistically significant correlations were found for $d_u^{\rm tone, conf}$, certain trends were observed. 
Specifically, a negative correlation for neuroticism was observed in the Humorous group ($r = -0.35$, adjusted $p = 0.085$), and a positive correlation for openness was found in the Formal group ($r = 0.36$, adjusted $p = 0.15$).

\subsubsection{Qualitative Analysis}\label{subsubsec:discussions3}

As with user experiments 1 and 2, we analyzed the open-ended responses (Q1 and Q2).

\paragraph{Impact of Explanations on Decision-making (Q1)}
A total of 24\% of participants responded that explanations did not affect or inform their decisions.
This percentage is approximately 10 points lower than that in other scenarios, suggesting that completing the task without explanation may have been more challenging.
Many participants who said they were influenced mentioned that they thought the AI had more expertise than they did and respected the AI's assessment, or that the AI provided reasonable justifications.
Conversely, those who said they were not influenced often noted that ``the AI's explanation was useful but lacked consideration of emotional aspects,'' ``it didn't provide new information,'' or ``I made my prediction based on my own perspective.''
These participants were likely less affected by the tone of the explanation.

\paragraph{Impact of Tone Changes on Decision-making (Q2)}
Among the 135 participants who noticed a change in tone and provided relevant responses, 59 participants said it did not affect their decisions, whereas 55 participants said it did.
Some participants felt that an informal tone was inappropriate for a serious topic, such as recidivism risk, whereas others found the tone easier to understand.
For example, P9, who was assigned the casual tone, said, ``The explanation felt more human and trustworthy,'' and P152 remarked, ``It felt like talking to a colleague, so it was easier to understand.''
Conversely, P99 felt it ``lacked persuasion,'' and P137 said, ``The tone was inappropriate for such a serious topic and distracted me overall.''
Interestingly, some participants felt that informal tones, such as casual or humorous, conveyed a sense of bias.
For example, P164 said, ``The tone was too light and gave the impression of bias.''
For the Authoritative group, P169 found it ``persuasive and logical,'' and P184 described it as ``very precise and calculated.''
However, P210 felt that ``the extreme language undermined the accuracy.''
These findings reveal that even with the same tone, participants' impressions of the AI varied.

\section{Discussion}\label{sec:discussion}
We investigated how the explanation tone impacts user decision-making across three scenarios with different AI roles: movie recommendation (assistant), opinion formation (second-opinion provider), and recidivism risk prediction (expert).
Our study revealed that tone has a significant overall effect in the opinion formation scenario and that certain user attributes demonstrate sensitivity to tone in other scenarios.
This section summarizes the findings, discusses their generalizability, addresses concerns and countermeasures related to tone adjustment, and outlines the limitations of this study.

\begin{table*}[t]
\caption{Summary of the results for each scenario. The \emph{Overall} column indicates whether there was a statistically significant overall effect across participants, independent of user attributes, while the columns for each user attribute indicate whether the tone and the respective attribute manifested interaction effects.
Checkmarks $\checkmark$ represent statistically significant differences based on a (two-way) ANOVA.
Tone pairs with statistically significant differences identified through post-hoc tests, as well as tones with statistically significant correlations from correlation tests, are shown in bold.
Non-bold items represent trends approaching significance.
Additionally, $+$ and $-$ symbols indicate positive and negative correlations, respectively.} 
\label{tb:summary}
\fontsize{6.5pt}{9pt}\selectfont
\centering
{\small \textbf{Recommendation Scenario}}\\
\begin{tabular}
{@{}p{35pt}|p{30pt}|p{35pt}p{45pt}p{40pt}p{40pt}p{48pt}p{40pt}p{45pt}}
\toprule
Metric & Overall & Gender & Age & Extroversion & Agreeableness & Conscientiousness & Neuroticism & Openness \\ \midrule
$d_u^{\rm tone}$ &  & \textbf{$\checkmark$} & \textbf{Formal (+)} &  & Formal (+) & Neutral (-) &  & Formal (+) \\ 
 &  & (Romantic) & \textbf{Romantic (-)}, &  &  &  &  & \\ 
  &  &  & Humorous (-) &  &  &  &  & \\ 
\bottomrule
\end{tabular} \\ 

\vspace{3pt}

{\small \textbf{Opinion Formation Scenario}}\\
\begin{tabular}
{@{}p{35pt}|p{30pt}|p{35pt}p{45pt}p{40pt}p{40pt}p{48pt}p{40pt}p{45pt}}
\toprule
Metric & Overall & Gender & Age & Extroversion & Agreeableness & Conscientiousness & Neuroticism & Openness \\ \midrule
$d_u^{\rm tone, stance}$ & \textbf{$\checkmark$} &  &  &  &  &  &  &  \\ 
 & \textbf{F. vs. H.} &  &  &  &  &  &  &  \\ 
  & \textbf{H. vs. N.} &  &  &  &  &  &  &  \\ 
$d_u^{\rm tone, conf}$ & \textbf{$\checkmark$} &  & Authoritative (+) &  &  &  &  & Authoritative (+) \\ 
 & F. vs. H. &  &  &  &  &  &  & Casual (+) \\ 
$d_u^{\rm tone, know}$ &  &  & Authoritative (+) &  &  &  &  &  \\ 
\bottomrule
\end{tabular}
\\

\vspace{3pt}

{\small \textbf{Advice Scenario}}\\
\begin{tabular}
{@{}p{35pt}|p{30pt}|p{35pt}p{45pt}p{40pt}p{40pt}p{48pt}p{40pt}p{45pt}}
\toprule
Metric & Overall & Gender & Age & Extroversion & Agreeableness & Conscientiousness & Neuroticism & Openness \\ \midrule
$d_u^{\rm tone, risk}$ &  & \textbf{$\checkmark$} & Formal (-) &  & Humorous (-) & Humorous (-) &  & Humorous (-) \\ 
 &  & (Authoritative) & Neutral (-) &  &  &  &  & \\ 
$d_u^{\rm tone, conf}$ & \textbf{$\checkmark$}  & \textbf{$\checkmark$} & Authoritative (+) &  &  &  & Humorous (-) & Formal (+) \\ 
 & C. vs. N. & (Formal) &  &  &  &  &  &  \\ 
\bottomrule
\end{tabular}
\end{table*}

\subsection{Summary of Findings and Interpretations}\label{subsec:discussion_summay}

Table~\ref{tb:summary} summarizes our results.
The \emph{Overall} column indicates whether the tone had an overall effect, whereas other columns indicate interactions with specific attributes.
Bold tones represent statistically significant based on post-hoc tests or non-correlation tests, while others reflect trends approaching significance.
The following subsections discuss the results based on scenario characteristics and user attributes.

\subsubsection{Scenario Characteristics (Overall Effect Irrespective of User Attributes)}

In the opinion formation scenario, the humorous tone had a general effect regardless of user attributes.
In contrast, the tone had no significant overall effect in the movie recommendation and recidivism prediction scenarios, though its impact varied by user attributes.
These differences likely stem from task characteristics, AI roles, participants' expectations for the explanations, and how the explanations were utilized.

In the opinion formation scenario, a neutral tone in Phase 2 reinforced participants' supportive stance (Fig.~\ref{fig:opinion_agreement_change_1to2}), while a humorous tone in Phase 3 caused a significant shift in the opposite direction. 
This was probably because the humorous tone lacked seriousness, undermining its credibility and eliciting resistance. 
This effect was consistent across all participants, underscoring the role of tone in shaping decision-making alongside personal values.

On the other hand, in the movie recommendation scenario, tone had no significant overall effect, likely due to three factors.
First, participants prioritized content over tone when assessing recommendation alignment, consistent with previous findings~\cite{okoso2024impact} that tone does not significantly impact users' perceptions of the system (e.g., persuasiveness or trustworthiness) for movie recommendations. 
Second, differences in the participants' expectations regarding explanations may have contributed to limiting the effect. Some participants reported explanations in Phase 2 insufficient and relied more on movie plots or genres, as reflected in minimal score shifts ($d_u^{\rm exp} \approx 0$, Fig.\ref{fig:movie_score_change_1to2}).
These participants seemingly placed little value in the explanations, which could explain the limited impact of tone.
Third, algorithm aversion~\cite{mahmud2022influences} may have played a role. Some participants expressed discomfort with AI mimicking human emotions, potentially reducing tone’s influence.

In the advice scenario, explanations played a crucial role as supplementary information (See Sec.~\ref{subsubsec:discussions3} and Fig.~\ref{fig:advice_pred_score_abs_1to2}).
Interestingly, questionnaire responses revealed three distinct approaches to using explanations, which may explain the limited overall effect of tone.
First, some participants followed a \emph{sequential paradigm}~\cite{tejeda2022ai}, initially making their own predictions and then comparing them with AI’s predictions.
If no discrepancies were perceived, they likely disregarded the explanations. 
This behavior aligns with previous research~\cite{tran2023user}, which suggests that users primarily need explanations when AI recommendations in high-stakes domains are unsatisfactory.
Repeated absence of discrepancies may also lead participants to fully rely on AI predictions, exhibiting automation bias~\cite{bertrand2022cognitive}.
Second, another group followed a \emph{simultaneous paradigm}~\cite{tejeda2022ai}, first reviewing AI predictions, then adjusting risk scores by integrating profile data and explanations while considering their opinions and emotions.
These participants may have been emotionally influenced by the tone, which consequently shaped their perception of the information’s trustworthiness.
Third, some participants exhibited algorithm aversion~\cite{mahmud2022influences}, relying solely on profile data and largely ignoring both AI predictions and explanations, resulting in minimal influence from tone.
Thus, the differences in decision-making approaches stemming from the task characteristics likely influenced the effect of tone in the recidivism prediction scenario where AI acted as an expert.

In summary, the differing effects of tone across scenarios with different AI roles likely stemmed from users’ varying expectations for explanations and their different ways of utilizing them.
When AI acted as a second-opinion provider (opinion formation scenario), users tended to view explanations as reinforcing their beliefs or offering new perspectives, using tone to assess their trustworthiness.
In scenarios where AI served as an assistant (movie recommendation scenario), users focused on content to determine whether recommendations aligned with their preferences, often disregarding explanations that did not meet their expectations.
When AI acted as an expert (recidivism prediction scenario), users treated the AI’s suggestions and explanations as essential information, with the impact of tone likely varying depending on how they referenced explanations (sequential vs. simultaneous paradigm).

\subsubsection{User Attributes}

Age showed the most significant interaction with tone.
In the recommendation scenario, older participants tended to rate movies lower with humorous or romantic tones but higher with a formal tone.
Similarly, in the recidivism prediction scenario, older participants showed smaller prediction changes with formal or neutral tones, suggesting greater a reliance on AI advice.
Additionally, as age increased, authoritative tones were found to enhance confidence and perceived knowledge, potentially reinforcing the illusion of explanatory depth~\cite{rozenblit2002misunderstood}.
These findings suggest that older participants were unconsciously influenced by tone in their decision-making.

Personality traits interacted most notably with tone in the advice scenario.
Participants with high agreeableness were more receptive to AI suggestions in a humorous tone, likely due to their tendency to avoid conflict.
Participants with low conscientiousness tend to act intuitively and may perceive a humorous tone as inappropriate or unprofessional for serious tasks such as recidivism prediction, leading to negative reactions.
Similarly, participants with low openness often resist change, which may have caused them to feel uncomfortable with a humorous tone and respond negatively.

Notably, we found no interaction between extroversion and tone across all scenarios, which contrasts with the findings of previous research~\cite{okoso2024impact}.
While extroversion has a positive impact on users' perceptions, such as persuasiveness and trust, in recommendation scenarios, our results suggest it does not significantly affect decision-making. 
This discrepancy may be because extroverted individuals tend to adapt well to system interactions, resulting in positive perceptions, while asserting their opinions strongly, thereby limiting how much tone influences their final decisions.
These findings highlight the importance of investigating both perceptual outcomes and the direct impact of tone on decision-making.

\subsection{Generalizability}\label{subsec:discussion_generalizability}

This study reveals that the tone of explanation can influence users' decision-making in scenarios where AI performs as an assistant, second-opinion provider, or expert, and our findings can be applied to other scenarios in which AI has similar roles.
In the opinion formation scenario, AI provided a second opinion, and a significant tone effect was observed, independent of user attributes.
The users appeared to rely on tone as a trust-calibration~\cite{zhang2020effect} tool to assess the second opinion’s credibility.
Similar effects are expected in other scenarios where AI provides a second opinion, such as investment decisions, health consultations, and career advice.
In the movie recommendation scenario where AI functioned as an assistant, users prioritized the content of the explanations, and the effects of tone tended to vary with user attributes.
Similar outcomes are expected regarding other assistant roles, such as shopping or travel planning, where explanations help users verify that the advice conforms to their preferences.

Generalizing scenarios in which AI functions as an expert, such as recidivism risk prediction, may depend on user expertise.
For expert users, AI may also serve as a second-opinion provider~\cite{sivaraman2023ignore}, potentially exhibiting effects similar to those in opinion formation scenarios.
For non-expert users, the outcomes may differ depending on whether the task has personal or extra-personal importance.
In tasks such as predicting recidivism, where outcomes have extra-personal effects, explanations may be overlooked owing to a sequential paradigm~\cite{tejeda2022ai} or automation bias~\cite{bertrand2022cognitive}, which limits tone effects.
Similar trends may occur in scenarios such as loan repayment or income prediction~\cite{lai2023towards}, where AI provides expert input.
However, in tasks such as medical diagnoses or personal loan approvals, when the user is directly affected, explanations are less likely to be ignored.
In such cases, tone could be crucial in trust calibration.

In addition, our results suggest that older users prefer formal tones and are more likely to follow AI's suggestions when delivered in a formal tone.
However, other research~\cite{chin2024like} focusing on voice assistants has reported that older users prefer casual tones over formal ones, which contrasts with our findings.
This discrepancy suggests that the mode of interaction with AI may influence user decision-making.
While our study assumed text-based interactions, different outcomes might be observed in the context of voice-based interactions.

\subsection{Expectations, Concerns, and Countermeasures Related to Tone Adjustments}\label{subsec:discussion_conserns}

In our user experiments, we analyzed user expectations and concerns regarding AI-adjusted explanation tones through responses to an open-ended question (Q3).
To categorize the responses, we labeled them with one or more of the following categories: expectation, concern, impression (positive, negative, neutral), opinion, or other.
Two of our authors independently labeled the data and disagreements were resolved through discussion.

Expectations for AI-driven tone adjustment included improved information clarity, enhanced user experience, diverse perspectives, and tone personalization.
A total of 39 participants expected tone adjustment to clarify information, simplify complex tasks, and facilitate better decision-making.
Another 35 participants believed it could foster empathy with the AI system, make recommendations more appealing, and enhance the overall user experience.
One participant (recommendation scenario: P193) mentioned that beautifying boring text by the choice of tone could make explanations more engaging and increase interest in the suggested movies.
Additionally, 10 participants viewed tone adjustment positively for its ability to highlight different aspects of the text, helping them notice overlooked details and think from multiple perspectives.
Furthermore, 16 participants stated that allowing users to choose tones based on personal preferences would further enhance the user experience.

However, several concerns about tone adjustment were highlighted, including intentional manipulation or bias, confusion from inconsistent or inappropriate tones, and reduced trust in the AI system.
The most frequently mentioned concern was intentional manipulation or bias, raised by 61 participants.
Specific concerns included that tone adjustment engendered a sense that the AI was trying to manipulate human decision-making (recommendation scenario: P271), that the AI might appear to be biased (advice scenario: P118), and that this could be misused to steer users in specific directions (advice scenario: P119).
In addition, 19 participants noted that sudden tone changes or inconsistency could confuse users, and 31 participants said that inappropriate tones could undermine the AI's credibility.
Furthermore, several participants found it unsettling that AI systems used human-like speech despite being emotionless.

To address these concerns, we propose three key approaches for designers of AI systems with explanatory capabilities.
First, the systems should allow users to view the rationale for tone settings and provide an option to disable tones by reverting to a neutral tone.
This promotes transparency and helps users understand that tones are chosen appropriately, thereby enhancing trust in the system.
Second, tone settings should align with the AI's role or users should be able to choose the tone.
For example, setting tones aligned with commonly understood roles, such as adopting a formal tone when an AI functions as an expert, can prevent the tone from creating an inappropriate impression.
If users can select tones, they should be informed about the potential impacts on their experience and decision-making.
Third, designers should undergo ethical training and implement audit mechanisms. 
Previous research~\cite{yang2024fair} has shown that guidance on fair decision-making can promote efforts to correct their biases.
Training and audits can help minimize the risks of inappropriate tone settings or unintentional biases.
Tone adjustments can enhance the user experience, but careful design and operation are crucial.
These approaches aim to preserve the benefits of tone adjustment while addressing ethical concerns and offering guidance for responsible implementation of AI systems with explanatory capabilities.

\subsection{Limitations}\label{subsec:limitations}

Our study has the following limitations:
\begin{itemize}
    \item Although we used six discrete tones in this study, the actual expression of explanations can be much more diverse and vary in intensity.
    Therefore, different results may be obtained if other forms of expression, including conceptual metaphors or emotional warmth, were used.
    \item There was a sample size imbalance across the intervention groups, which may have reduced the statistical power. This study estimated a sample size of 50 participants per group, assuming $80\%$ power, a $5\%$ significance level, and an effect size of $0.5$. However, after excluding participants who failed the attention check or whose open-ended responses were of notably low quality, certain groups did not reach this target. In particular, the sample size for the advice scenario was smaller than planned, which may have led to reduced statistical power and lower reliability and inconsistency in the results compared to other scenarios.
    \item Participants' preconceptions about AI-generated explanations may also have influenced the results. There were participants who had a neutral or positive perception of AI, as well as those who had a skeptical or negative preconception of AI. We found that the latter group largely performed the task without relying on the AI-generated explanations. These preconceptions may have partially obscured the tone effects in the explanations, potentially introducing bias into the results.
    \item This study did not account for the interrelationship between tone and task type. For example, in the movie recommendation scenario, the effect of tone may vary depending on the genre. Future research should explore the interaction between task characteristics and tone.
\end{itemize}
In light of these limitations, future research should examine the effects of tone across a wider range of contexts and expressions, while ensuring larger sample sizes and stricter control over the influence of participants' preconceptions. Additionally, it is important to investigate further the interaction between tone and the nature of the task.

\section{Conclusion}\label{sec:conclusion}

This study investigated how the tone of an explanation influences user decision-making by focusing on the roles of AI and user attributes. We evaluated three scenarios representing different AI roles: movie recommendation (assistants), opinion formation (second-opinion providers), and recidivism risk prediction (experts).
We generated datasets with explanations for various tones using LLMs.
Our user experiments revealed that the impact of tone varied depending on the AI's role, probably because of differences in user expectations for explanations and how explanations were referenced.
The tone effects also differed according to user attributes, with older users being more susceptible to being influenced by tone.
We also found that with highly extroverted users discrepancies occurred between their perceptions of the system and their decision-making.
Furthermore, open-ended responses indicated that, while tone adjustments can enhance clarity and the user experience, they also raised concerns about malicious manipulation, bias, confusion due to inconsistency in tone, and reduced trust arising from inappropriate tones.
Our findings highlight the risk that certain user groups are unintentionally affected by the tone of explanation.
This study provides crucial insights for designing explanations tailored to AI roles and offers guidance to the HCI community and practitioners on improving the user experience while addressing ethical considerations.

\bibliographystyle{ACM-Reference-Format}
\bibliography{references}

\appendix
\section{Dataset Generation}
We implemented the dataset-generation procedure in Python using the OpenAI API ~\footnote{https://openai.com/blog/openai-api}.
We used the DALL-E 3 model to generate poster images in the recommendation scenario and the GPT-4o model with the temperature set to a default value of $0.7$ for other generations.

\subsection{Recommendation Scenario}\label{app:dataset_recsys}

The process for generating information for fictional movie items and the corresponding advertising text in each tone is as follows:
\begin{enumerate}
    \item We generated $35$ fictional movie items, with five items for each of the seven genres. Each item included a title and a short plot of approximately $50$ words. The prompt used was as follows:
    \begin{itembox}[l]{Prompt: Generation of Movie Item }
    Generate five fictional movie titles and short plots (around $50$ words each) in the \texttt{[genre]} genre.
    \end{itembox}
    Here, \texttt{[genre]} represents each of the following seven genres: Action, Comedy, Drama, Documentary, Science Fiction, Thriller, and Romance.
    \item We then created advertisements of approximately $50$ words based on the generated movie items. The system was set with the role, ``You are a professional copywriter,'' and the following prompt was used:
    \begin{itembox}[l]{Prompt: Advertisement Text Generation}
    Generate a promotional blurb (around $50$ words) for a movie based on the following title and plot. Focus on one or more of the following aspects: plot, theme, genre, characters, direction, cast, director, music, and awards. Please use vague references or pseudonyms instead of actual names for directors, actors, and other individuals.\\
    \texttt{[movie item]}
    \end{itembox}
    Here, \texttt{[movie item]} includes the information for each movie item.
    \item The generated advertisement texts were converted to the specified tones. The content remained unchanged, and the rewritten texts were limited to approximately $50$ words. The system was set with the role, ``You are a professional copywriter specializing in the \texttt{[tone]} tone,'' and the following prompt was used:
    \begin{itembox}[l]{Prompt: Tone Conversion of Advertisement Text}
    Rewrite the following promotional blurb in a/an \texttt{[tone]} style. Ensure that the tone aligns with the defined characteristics: \texttt{[definition]}. The content and meaning of the original text should remain unchanged. Additionally, limit the revised text to around $50$ words.\\
    Original: \texttt{[original text]}
    \end{itembox}
    Here, \texttt{[tone]} includes each of the target tones: neutral, formal, humorous, authoritative, or romantic. Moreover, \texttt{[definition]} includes the definition of each tone, as listed in Table~\ref{tb:tone_definitions}, and \texttt{[original text]} represents the original advertisement text.
    \item We generated poster images based on the information of the movie item. The size was set to $1024\times1792$, and the following prompt was used:
    \begin{itembox}[l]{Prompt: Poster Image Generation}
    Create a movie poster for the following movie:\\
    Original: \texttt{[movie item]}
    \end{itembox}
\end{enumerate}

\subsection{Opinion Formation Scenario}\label{app:dataset_opinion}

The generation procedure for the opinion dataset is as follows:
\begin{enumerate}
    \item For each topic, we generated two supportive opinions, each consisting of approximately 50 words. The system was assigned the role, ``You are an expert in the field,'' and the following prompt was used:
    \begin{itembox}[l]{Prompt: Generation of Opinions}
    Generate two supportive opinions on the topic. Offer unique perspectives that highlight different benefits or positive aspects. Ensure each opinion is concise and around 50 words.\\
    \texttt{[topic]}
    \end{itembox}
    Here, \texttt{[topic]} represents the specific topic.
    We reviewed the generated opinions and selected one for use in the experiment.
    \item The generated opinions were converted to the specified tones. The content remained unchanged, and the revised opinions were kept at approximately 50 words. The system's role and the prompts used were essentially the same as those in the recommendation scenario, with slight modifications to fit the opinion formation context. The \texttt{[tone]} in the prompt was set to each of the following: neutral, formal, authoritative, casual, or humorous.
\end{enumerate}

\subsection{Advice Scenario}\label{app:dataset_advice}

The procedure for generating the advice dataset is as follows:
\begin{enumerate}
    \item Based on the profile information, we generated predicted recidivism risk scores and the corresponding reasoning. The system was set with the role ``You are an expert in risk assessment,'' and the following prompt was used:
    \begin{itembox}[l]{Prompt: Advice Generation}
    Predict the recidivism risk score for the following person. The risk score should reflect the likelihood of this person committing another crime within 2 years. The score should be between 1 and 10, with 10 indicating the highest risk. Additionally, provide a brief explanation (approximately 50 words) of why this score was given.\\
    Profile: \texttt{[profile]}
    \end{itembox}
    Here, \texttt{[profile]} represents the profile information of each defendant. The prediction results showed an accuracy of $54.0\%$ for the recidivism risk labels and a mean absolute error of $2.17$ for the predicted risk scores. 
    These evaluations refer to the comparison with the COMPAS algorithm's assessment, not actual recidivism outcomes. 
    We randomly selected 30 of the 54 defendants whose predicted risk labels matched their COMPAS labels.
    For this subset, the average absolute error of the predicted risk score was $1.27$.
    \item The generated reasoning were converted to the specified tones. The content remained unchanged, and the revised texts were kept to approximately 50 words each. The system's role and the prompt used were essentially the same as those in the recommendation and opinion formation scenarios, with slight adjustments to fit the advice scenario. The \texttt{[tone]} in the prompt was set to each of the following: neutral, formal, authoritative, casual, or humorous.
\end{enumerate}

\begin{figure*}[t]
    \centering
    \begin{minipage}[tb]{\linewidth}
        \centering
        \includegraphics[width=0.85\linewidth]{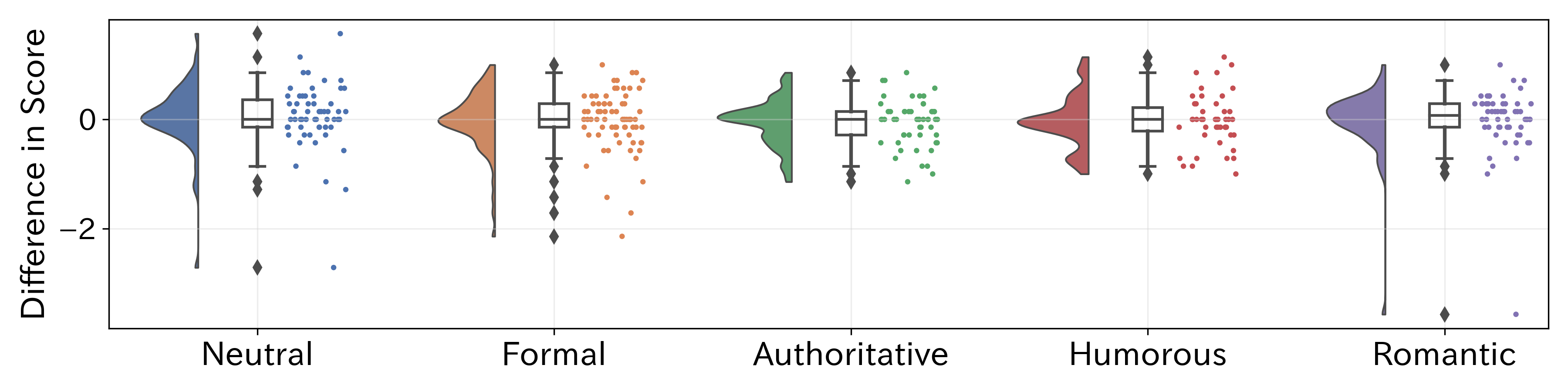}
        \subcaption[]{Recommendation scenario ($d_u^{\rm exp}$).}
        \label{fig:movie_score_change_1to2}
    \end{minipage}
    \vfill
    \begin{minipage}[tb]{\linewidth}
        \centering
        \includegraphics[width=0.85\linewidth]{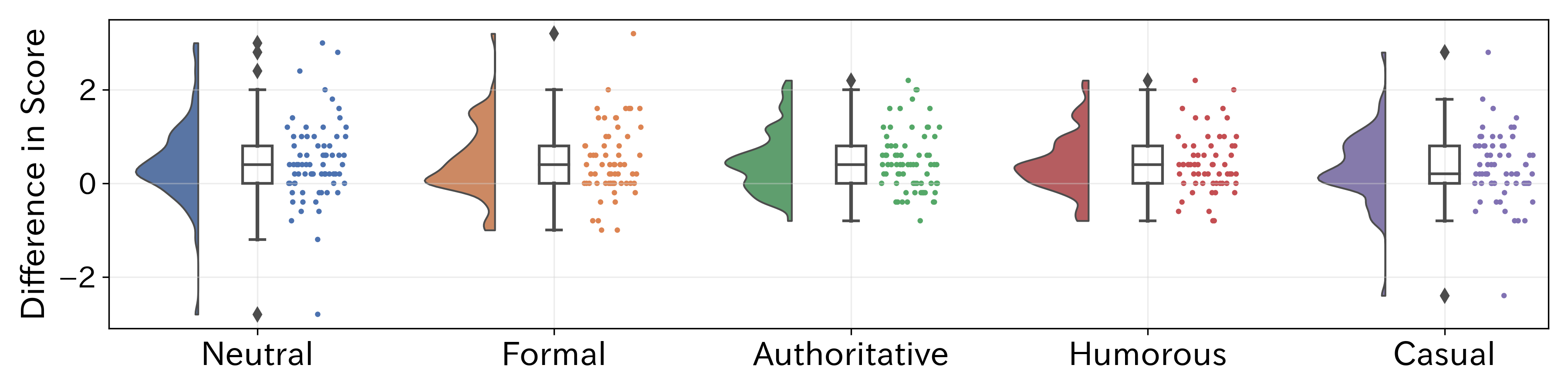}
        \subcaption[]{Opinion formation scenario ($d_u^{\rm exp, stance}$).}
        \label{fig:opinion_agreement_change_1to2}
    \end{minipage}
    \vfill
    \begin{minipage}[tb]{\linewidth}
        \centering
        \includegraphics[width=0.85\linewidth]{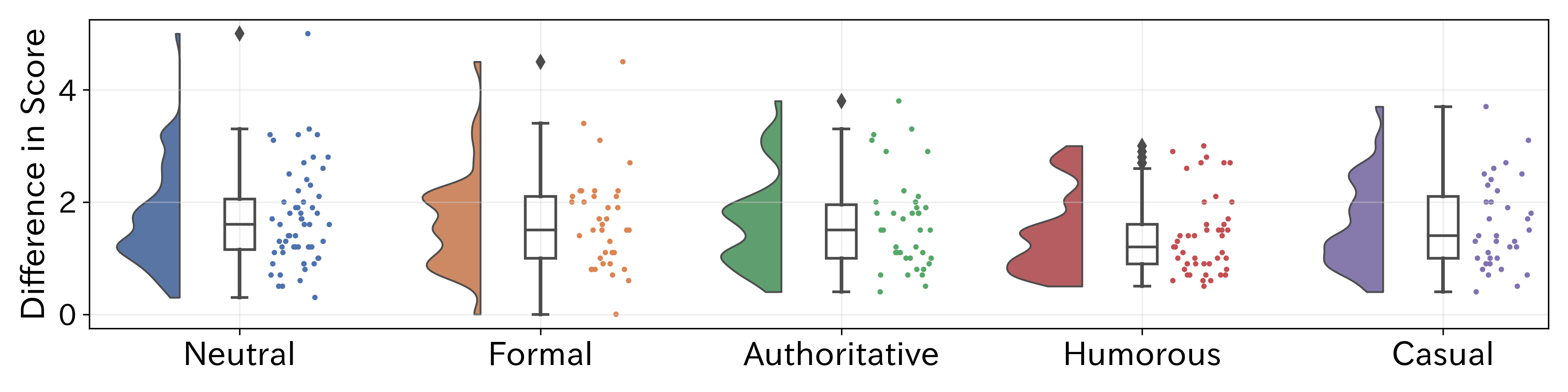}
        \subcaption[]{Advice scenario ($d_u^{\rm exp, risk}$).}
        \label{fig:advice_pred_score_abs_1to2}
    \end{minipage}
    \caption{Distributions of score differences by explanation (Phase 1 to 2) for each experiment scenario.}
    \label{fig:distritbutions_1to2}
    \Description{The figure contains three subfigures, each representing the distribution of score differences across five tone interventions (Neutral, Formal, Authoritative, Humorous, and Casual) in each scenario (recommendation, opinion formation, advice). Each subfigure uses violin and box plots to depict the range and concentration of score differences for a specific metric.}
\end{figure*}

\section{Distribution of Each Score among Intervention Group (Phases 1 to 2.)}
Figure~\ref{fig:distritbutions_1to2} illustrates the distributions of the score differences ($d_u^{\rm exp}$, $d_u^{\rm exp, stance}$, $d_u^{\rm exp, risk}$) across the intervention tone groups in each scenario.
In the recommendation scenario, the average score difference was close to zero, indicating that, on average, the presence or absence of explanations did not result in significant score changes.
In contrast, in the opinion formation scenario, the average score difference was greater than zero, suggesting that the participants were influenced by the AI's supportive explanations of the topics.
Additionally, in the advice scenario, the absolute value of the average score difference exceeded one, indicating that the participants likely relied on the AI's explanations to complete the tasks.

\end{document}